\documentclass[lettersize,journal]{IEEEtran}
\usepackage{amsmath,amsfonts}
\usepackage{algorithmic}
\usepackage{algorithm}
\usepackage{array}
\usepackage{textcomp}
\usepackage{stfloats}
\usepackage{url}
\usepackage{verbatim}
\usepackage{graphicx}
\usepackage{amssymb}
\usepackage{amsfonts}
\usepackage{amsmath}
\usepackage{physics}
\usepackage{algorithmic}
\usepackage{algorithm} 
\usepackage{graphicx}
\usepackage{color}
\usepackage{bm}
\usepackage{epstopdf} 
\usepackage{subfigure}
\usepackage{booktabs}
\usepackage[dvipsnames]{xcolor}
\usepackage{cite} % 多项引用
\usepackage{balance}
\usepackage{setspace}  % 改变行距
\usepackage{amsthm}
\usepackage{amsmath} 
\usepackage{hyperref}
\newcounter{MYtempeqncnt}
\newtheoremstyle{note}
{3pt}%
{3pt}%
{}%
{\parindent}%
{\rmfamily \bfseries}%
{:}%
{5pt}%
{}%

\theoremstyle{note}

\newtheorem{ppn}{Proposition}
\newtheorem{lemma}{Lemma} 
\newtheorem{remark}{Remark} 

\newtheorem*{pf}{Proof}

\newcommand{\lmref}[1]{\textbf{Lemma \ref{#1}}}
\newcommand{\ppnref}[1]{\textbf{Proposition \ref{#1}}}
\newcommand{\remarkref}[1]{\textbf{Remark \ref{#1}}}
\newcommand{\algref}[1]{\textbf{Algorithm \ref{#1}}}

\newcommand{\figref}[1]{Fig. \ref{#1}}

\begin{document}
	
	\title{Robust Symbol-Level Precoding for Massive MIMO Communication Under Channel Aging}
	
	\author{Yafei Wang, \textit{Graduate Student Member}, \textit{IEEE}, Xinping Yi, \textit{Member}, \textit{IEEE}, \\ Hongwei Hou, \textit{Graduate Student Member}, \textit{IEEE},
	Wenjin Wang, \textit{Member}, \textit{IEEE}, Shi Jin, \textit{Fellow}, \textit{IEEE}% <-this % stops a space
	\thanks{Manuscript revised xxx.}
		\thanks{Yafei Wang, Hongwei Hou, Wenjin Wang, and Shi Jin are with the National Mobile Communications Research Laboratory, Southeast University, Nanjing 210096, China (e-mail: wangyf@seu.edu.cn; 	
		hongweihou@seu.edu.cn; wangwj@seu.edu.cn; jinshi@seu.edu.cn).}
		\thanks{Xinping Yi is with the Department of Electrical Engineering and Electronics, University of Liverpool, Liverpool L69 3BX, U.K. (e-mail: xinping.yi@liverpool.ac.uk).}}
		
	\markboth{}%
	{Shell \MakeLowercase{\textit{et al.}}: A Sample Article Using IEEEtran.cls for IEEE Journals}
	
	\maketitle

	\begin{abstract}
		This paper investigates the robust design of symbol-level precoding (SLP) for multiuser multiple-input multiple-output (MIMO) downlink transmission with imperfect channel state information (CSI) caused by channel aging. By utilizing the \textit{a posteriori} channel model based on the widely adopted jointly correlated channel model,
		the imperfect CSI is modeled as the statistical CSI incorporating the channel mean and channel variance information with spatial correlation. With the signal model in the presence of channel aging, we formulate the signal-to-noise-plus-interference ratio (SINR) balancing and minimum mean square error (MMSE) problems for robust SLP design. The former targets to maximize the minimum SINR across users, while the latter minimizes the mean square error between the received signal and the target constellation point. When it comes to massive MIMO scenarios, the increment in the number of antennas poses a computational complexity challenge, limiting the deployment of SLP schemes. To address such a challenge, we simplify the objective function of the SINR balancing problem and further derive a closed-form SLP scheme. Besides, by approximating the matrix involved in the computation, we modify the proposed algorithm and develop an MMSE-based SLP scheme with lower computation complexity. Simulation results confirm the superiority of the proposed schemes over the state-of-the-art SLP schemes.
	\end{abstract}
	
	% Note that keywords are not normally used for peerreview papers.
	
	\IEEEpeerreviewmaketitle
	
	\section{Introduction}
	\IEEEPARstart{A}{s} 
	a widely adopted technology in multiuser multi-input multi-output (MU-MIMO) transmission,  precoding is used to exploit potential capacity gains by alleviating user interference. Linear precoding schemes, including the zero-force (ZF) precoding \cite{1040325} and the minimum mean square error (MMSE) precoding \cite{9054488}, design the precoding matrix using channel state information (CSI). Such schemes usually require low computational complexity {at the cost of suboptimal performance}. Compared to linear precoding, nonlinear precoding, {e.g., symbol-level precoding (SLP) \cite{8359237, 9035662, Li2021,4801492,5605266,6875291,7103338,7417066,8462190,7942010,8575164,8466792,8299553,8477154,8815429,8374931,9120670,shao2020minimum, mohammad2021unsupervised, 8647896, 9435988, hegde2019interference, 9025051},} exploits both CSI and input data to achieve superior performance.
	
	SLP optimizes the transmit signal at a symbol level by leveraging both CSI and constellations of the user symbols, with a key emphasis on interference exploitation \cite{8359237, 9035662, Li2021,4801492,5605266,6875291,7103338,7417066,8462190,7942010,8575164,8466792,8299553,8477154,8815429,8374931,9120670,shao2020minimum, mohammad2021unsupervised, 8647896, 9435988, hegde2019interference, 9025051}.  
	By maintaining the constructive interference (CI) and suppressing the destructive interference (DI), ZF and regularized ZF linear precoding were modified to be SLP schemes in \cite{4801492}, while the concept {of} converting DI to CI was further considered in \cite{5605266}. The CI constraint was introduced to confine the noise-free received signal within the CI region (CIR), where the interference component of the received signal was CI \cite{7103338}. Based on the CI constraint for phase-shift keying (PSK) and quadrature amplitude modulation (QAM) symbols, power minimization (PM) and signal-to-noise-plus-interference ratio (SINR) balancing problems were widely investigated for SLP design \cite{6875291,7103338,7417066,8462190,7942010,8299553,8575164,8466792,Li2021,8477154,7472286,8445846,7042789}.
	The PM problem minimized the transmit power given the SINR requirement \cite{6875291,7103338,7417066,8462190,7942010,8299553,8477154,7042789,8575164}, and its optimal solution could be scaled to that of the SINR balancing problem \cite{7042789}, which targeted to maximize the minimum SINR while satisfying the transmit power constraint \cite{7103338,8466792,Li2021,8477154,7472286,8445846}. Moreover, \cite{9910472} investigated the MMSE problem that minimizes the mean square error (MSE) between the received signal and the expected signal located in CIR, based on which an SLP scheme was proposed to achieve significant performance with low complexity. Furthermore, the SINR balancing and MMSE-based SLP schemes have been proven to be the generalizations of ZF precoding and MMSE precoding \cite{8466792,Li2021,9910472}, respectively.
	
	Nevertheless, perfect CSI is required for most precoding schemes, where its acquisition at the base station (BS) is challenging in practical systems owing to various factors such as channel estimation error \cite{9449854}, phase noise drift \cite{9419749}, and channel aging \cite{9416909}. Therefore, in practice, it is urgently needed to design robust SLP against such CSI imperfection \cite{7103338, mohammad2021unsupervised, 8647896, 9435988, hegde2019interference, 9025051}.
	With the assumption of norm-bounded CSI error, the PM problem for SLP transmission was investigated in \cite{7103338}, following which an unsupervised deep unfolding framework was proposed to reach near-optimal performance with lower complexity in \cite{mohammad2021unsupervised}. Such an error model was also employed to design an SLP with probabilistic constraints in \cite{8647896} and extended to the SLP system with an intelligent reflecting surface in \cite{9435988}.
	Unlike the above fully digital SLP, a robust hybrid SLP was considered in \cite{hegde2019interference}, where the CSI error was modeled as phase errors associated with the analog precoder. Assuming an uncorrelated complex Gaussian additive error at the available CSI, the SINR-constrained SLP that could effectively improve energy efficiency was proposed in \cite{9025051}.
	
	As one of the most important reasons for CSI error in practical systems, the channel aging effect cannot be accurately captured by the existing CSI error models, such as norm-bounded or uncorrelated Gaussian additive models. The CSI errors due to channel aging are required to represent the variations of time-varying channels \cite{6608213,9416909,7307172}.
	The jointly correlated channel model is a widely adopted model to represent the spatial correlations of the channels \cite{1576533}, by which the \textit{a posteriori} channel model can be used to represent the channel under channel aging \cite{9516008,9592715,8694866,9745311}. Furthermore, unlike conventional robust precoding schemes \cite{8694866,7045498}, the SLP scheme that exploits both imperfect CSI and user symbols to combat channel aging at a symbol level has never been investigated.
	Thus, a question is raised: \textit{How to design robust SLP against channel aging based on the \textit{a posteriori} channel model?} This is the key issue to be investigated in this paper. By virtue of the \textit{a posteriori} channel model, the signal model under channel aging is formulated, and the robust SLP schemes are designed according to the carefully formulated SINR balancing and MMSE problems.
	In summary, our major contributions are as follows.
	\begin{itemize}
		\item Building upon the \textit{a posteriori} channel model, we construct the signal model of the downlink SLP transmission with channel aging. In this model, channel uncertainty introduces an extra Gaussian distributed interference term with power determined by the transmit signal. This model can also approximately represent the quasi-static scenarios.
		\item On the basis of the received signal model, we formulate the SINR balancing problem for robust SLP design, with channel uncertainty considered in the SINR representation. This problem is transformed into a typical max-min fractional programming (MMFP), which can be solved using the Generalized Dinkelbach's (GD) Algorithm \cite{zappone2015energy}. 
		In the case of massive MIMO, the above problem is simplified
		with an approximated objective function, for which we further propose a low-complexity SLP scheme with a closed-form solution. The simulation results show that the proposed two schemes greatly improve the minimum SINR and provide superior symbol error rate (SER) performance. 
		\item We formulate the MMSE problem for robust SLP design, which can suppress the interference caused by channel aging. {Given the problem structure, we derive an} alternating optimization algorithm for this problem. In the scenario of massive MIMO, we develop an MMSE-based SLP scheme that requires lower computational complexity by introducing the approximation of the matrix involved in the proposed algorithm. Simulations demonstrate that the above two schemes significantly minimize the MSE between the received signal and the target constellation, as well as decrease the SER during transmission.
	\end{itemize}
	
	The rest of the article is organized as follows: The signal model is built in Section \ref{Section 2}. The SINR balancing and MMSE problems with their corresponding robust SLP schemes are investigated in Section \ref{robust SB section} and Section \ref{robust MMSE section}, respectively. Simulation results are provided in Section \ref{Section 6}, and Section \ref{Section 7} concludes this article.
	
	{\textit{Notation}}: $x, {\bf x}, {\bf X}$ represent scalar, column vector, matrix. $(\cdot)^T$, $(\cdot)^{*}$, $(\cdot)^H$, and $(\cdot)^{-1}$ respectively denote the transpose, conjugate, transpose-conjugate, and inverse operations. ${\bf I}_{M}$ represents $M\times M$ identity matrix. $\left \|\cdot\right \|_{2}$ denotes $l_2$-norm. $\otimes$ and $\odot$ are the Kronecker
	product and Hardmard product operations. The operator ${\rm tr}(\cdot)$ represents the matrix trace.  ${{\rm diag}\{{\bf a}\}}$ represents the diagonal matrix whose diagonal elements are composed of ${\bf a}$. $\real(\cdot)$ and $\imaginary(\cdot)$ denote the real and imaginary parts of a complex scalar, vector, or matrix. $[{\bf X}]_{i,j}$ and $[{\bf x}]_i$ denotes the $(i,j)$-th and $i$-th element of ${\bf x}$ and ${\bf X}$, respectively. ${\bf x}{\succeq}{\bf 0}$ means all the elements of ${\bf x}$ is nonnegative. The expression $\mathcal{C}\mathcal{N}(\mu, \sigma^2)$ denotes circularly symmetric Gaussian distribution with expectation $\mu$ and variance $\sigma^2$. ${\mathbb{R}}^{M\times N}$ and ${\mathbb{C}}^{M\times N}$ represent the set of $M\times N$ dimension real- and complex-valued matrixes. $\nabla f$ denotes gradient of function $f(\cdot)$. $k\in \mathcal{K}$ means element $k$ belongs to set $\mathcal{K}$.
	
	\section{Signal Model}\label{Section 2}
	\subsection{System Model}\label{system model}
	
	Consider a MIMO downlink system operating in time division duplexing (TDD) mode, where a base station (BS) equipped with $N$-antenna uniform linear array (ULA) or uniform planar array (UPA) transmits the signal to $K$ single-antenna user equipment (UE) simultaneously in the same time-frequency resource. The time resources are divided into slots, where each time slot contains some uplink training symbols and $N_{\rm ds}$ downlink transmission symbols.
	The CSI is estimated from the uplink training symbols and utilized for the downlink transmission symbols.

	We use ${\bf h}_{k,n}\in{\mathbb{C}}^{N\times 1}$, $n=1,...,N_{\rm{ds}}$ to represent the channel from the BS to the $k$-th UE at $n$-th downlink transmission symbols, which is considered to be described by the widely adopted jointly correlated channel model \cite{9910031, 1576533, 5165196}. Define ${\bf h}^{\rm{u}}_{k,0}$ as the channel estimated at BS from the uplink training sequences in this slot by exploiting the channel reciprocity, and channel $ {\bf h}_{k, n}$ can be represented as the \textit{a posteriori} channel model \cite{8694866}:
	\begin{align}
		\begin{split}
			{\bf h}_{k, n} &= \alpha_{k,n} {\bf h}_{k,0}^{\rm{u}}+ \beta_{k,n}{\bf V}_{\rm D}^{*} ( {\bf m}_{k}\odot {\bf g}_{k,n}).
		\end{split}
		\label{posteriori}
	\end{align}
	In this formulation, ${\bf V}_{\rm D}\in{\mathbb{C}}^{N\times F_{\rm vh}N}$ is a matrix composed of (partial) discrete Fourier transform (DFT) matrix, where $F_{\rm vh}\in{\mathbb{N}}^{+}$ is the fine factor utilized to improve the fineness of the model \cite{9910031}; ${\bf g}_{k, n}\in{\mathbb{C}}^{F_{\rm vh}N \times 1}$ is a complex Gaussian random vector whose elements follow $\mathcal{C}\mathcal{N}(0, 1)$ independently; $ {\bf m}_{k}\in{\mathbb{R}}^{F_{\rm vh}N \times 1}$ is a sparse vector with nonnegative elements that remain constant for a relatively long period \cite{9516008,1576533}; The time variation of the
	channel is modeled by the first order Gauss-Markov process with the correlation coefficients $\alpha_{k,n}$ and ${\beta}_{k,n}=\sqrt{1-\alpha_{k,n}^{2}}$ that are related to the UE speed \cite{7307172,6608213}. Specifically, $\alpha_{k,n}$ is described by Jakes’ autocorrelation model \cite{9416909, 4299616}, i.e., $\alpha_{k,n} = J_0\left(2\pi v_k f_c nT/c\right)$, where $J_0(\cdot)$, $v_k$, $f_c$, $T$, and $c$ represent the first kind of Bessel functions of zero order, the speed of $k$-th UE, carrier frequency, the duration of a symbol, and the speed of light, respectively.
	
	For brevity, we temporarily drop the index $n$. At certain downlink symbol duration, the received signal of the $k$-th UE is
	\begin{equation}
		{y}_{k} = {\bf h}_{k}^T{\bf x}_{\rm c} + n_{k},\;\forall k \in {\mathcal{ K}},
		\label{received signal 1}
	\end{equation}
	where ${\mathcal{ K}}=\left\{1, 2, ..., K\right\}$, $n_{k}$ denotes the additive noise at the $k$-th UE following $\mathcal{C}\mathcal{N}(0, \sigma^2)$, ${\bf x}_{\rm c}\in{\mathbb{C}}^{N\times 1}$ is the transmit signal computed by the symbol-level precoder ${\rm SLP}\left(\cdot\right)$, i.e., 
	\begin{align}
		{\bf x}_{\rm c} &= {\rm SLP}\left({\bf s}_{\rm c},\left\{{ {\bf h}_{k}^{\rm{u}}}, {\bf m}_{k}, \alpha_{k}\right\}^{K}_{k=1}, \sigma^2\right),
	\end{align}
	where $\left\{{ {\bf h}_{k}^{\rm{u}}}, {\bf m}_{k}, \alpha_{k}\right\}^{K}_{k=1}$ and $\sigma^2$ are assumed to be available at the BS \cite{8694866,9910031,9745311}, ${\bf s}_{\rm c}$ denotes the $M$-PSK ($M> 2$) symbols to be sent to UEs at this symbol duration, which is given by
	\begin{align}
		{\bf s}_{\rm c} = \left[s_{1},s_{2},..., s_{K}\right]^T,
	\end{align}
	where $s_{k}$ is the symbol desired by the $k$-th UE. The signal to be demodulated is
	\begin{align}
		\begin{split}
			{\tilde{y}_{k}} &= {y}_{k}/{\gamma_{k}} = {\bf h}_{k}^T{\bf x}_{\rm c}/{\gamma_{k}} + n_{k}/{\gamma_{k}},
		\end{split}
	\end{align}
	where ${\gamma_{k}}$ is the rescaling factor provided by the SLP scheme  \cite{li2020symbol,8602458,9887802}.

	\subsection{Received Signal Model with Imperfect CSI}
	In this subsection, we build a received signal model given the \textit{a posteriori} channel model \eqref{posteriori}. By defining ${\bar {\bf h}}_k = \alpha_{k} {\bf h}^{\rm u}_{k}$ and substituting \eqref{posteriori} into \eqref{received signal 1}, we have 
	\begin{align}
		{y}_{k} = {\bar{\bf h}}_{k}^{T}{\bf x}_{\rm c} + {\beta}_{k}({{\bf m}}^T_{k}\odot {\bf g}^{T}_{k})  {\bf V}_{\rm D}^{H}{\bf x}_{\rm c} + n_{k},
		\label{received signal 2}
	\end{align}
	where the second term is the random interference caused by the channel uncertainty and can be simplified as 
	\begin{align}
		{\beta}_{k}({{\bf m}}^T_{k}\odot {\bf g}^{T}_{k})  {\bf V}_{\rm D}^{H}{\bf x}_{\rm c}= ({\beta}_{k}{\bar{\bf V}}_k{\bf x}_{\rm c})^T{\bf g}_{k},
	\end{align}
	where ${\bar{\bf V}}_k=\begin{bmatrix}
		{{\bf m}}_{k}\odot{\bf v}_1 & \cdots & {{\bf m}}_{k}\odot{\bf v}_N
	\end{bmatrix}$, and ${\bf v}_i$ is the $i$-th row of ${\bf V}_{\rm D}^{H}$. As the randomness comes from the last two terms, we define ${\bar{n}}_k=({\beta}_{k}{\bar{\bf V}}_k{\bf x}_{\rm c})^T{\bf g}_{k} + n_{k}$, and the received signal in \eqref{received signal 2} can be further rewritten as 
	\begin{align}
		{y}_{k} = {\bar{\bf h}}_{k}^{T}{\bf x}_{\rm c} + {\bar{n}}_k,
		\label{equivalent received signal}
	\end{align}
	where ${\bar{n}}_k\sim\mathcal{C}\mathcal{N}(0, {\beta}^2_k\|{\bar{\bf V}}_{k}{\bf x}_{\rm c}\|_{2}^{2}+\sigma^2)$ can be easily verified since the elements of ${\bf g}_k$ follow $\mathcal{C}\mathcal{N}(0, 1)$ independently. {Note that the average power of ${\bar{n}}_k$ is related to ${\bf x}_{\rm c}$.} Besides, the signal model \eqref{equivalent received signal} with imperfect CSI has a similar form as \eqref{received signal 1}, {where} ${\bar{\bf h}}_{k}$ and ${\bar{n}}_k$ will become ${\bf h}_{k}$ and ${{n}}_k$ for the quasi-static scenario, i.e., the correlation coefficient $\alpha_k$ is very close to 1.
	
	\begin{figure}[t]
		\centering
		\includegraphics[width=2.3in]{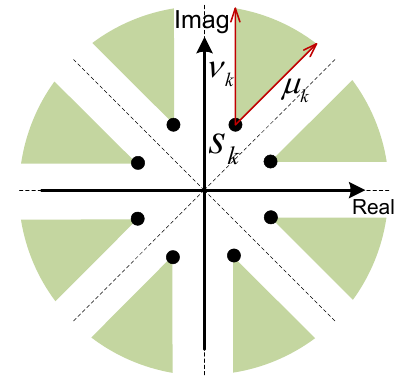}
		\caption{CIRs and their boundary vectors of 8PSK.}
		\label{CI_MMSE_CIR}
	\end{figure}
	\section{Robust Symbol-Level Precoding With SINR Balancing Criterion}\label{robust SB section}
	
	\subsection{Constructive Interference Region}
	
	CI is defined as the interference that moves the
	received signal further away from the decision boundaries to enhance  the demodulation performance, and CIR is the region where the signal is located with CI \cite{9910472, 8466792, Li2021}. Fig. \ref{CI_MMSE_CIR} shows the typical CIR (green areas) for 8PSK, where the dotted lines are the decision boundaries of $s_k$, ${{\mu}}_{k}$, and ${{\nu}}_{k}$ are boundary parameters of CIR corresponding to $s_k$. The typical CIR of $s_{k}$ can be expressed as \cite{9910472}
	\begin{align}
		\mathcal{D}_k = \left\{{{{\tilde {s}}_k}}|
		{{\tilde {s}}_k} = s_k + \delta_{\mu_k}{{\mu}}_{k}+\delta_{\nu_k}{{\nu}}_{k}
		,\ \delta_{\mu_k},\delta_{\nu_k}>0\right\},
		\label{CIR-D}
	\end{align}
	where ${{\mu}}_{k}$ and ${{\nu}}_{k}$ can be easily obtained from given $s_k$. 	
	
	\subsection{{SINR Balancing-Based Robust SLP}}\label{Section 4}
	SINR balancing aims to maximize the minimum SINR while satisfying the power constraint \cite{Li2021}. Based on the signal model \eqref{equivalent received signal}, we constrain ${\bar{\bf h}}^T_{k}{\bf x}_{\rm c}/{\gamma_{k}}$ (rescaled noise-free received signal) in the CIR, i.e., ${{\bar{\bf h}}^T_{k}{\bf x}_{\rm c}}/{\gamma_k}\in\mathcal{D}_{k}$,
	which can be further represented as \cite{8299553}
	\begin{align}
		{\bar{\bf h}}^T_{k}{\bf x}_{\rm c} \in \gamma_k \cdot\mathcal{D}_{k}, \;\forall k \in {\mathcal{ K}}.
		\label{CIR math}
	\end{align}
	The CI constraints convert the interference to CI, and all interference terms form components of the useful signal energy\cite{7103338,8477154}. Thus, according to \eqref{equivalent received signal}, \eqref{CIR-D}, and \eqref{CIR math}, the SINR of the $k$-th UE is given by
	\begin{align}
		{{\rm SINR}}_{k} = \frac{|{{\bar{\bf h}}}^T_k{\bf x}_{\rm c}|^2}{{\beta}^2_k\left \|{\bar{\bf V}}_{k}{\bf x}_{\rm c}\right\|_{2}^{2}+\sigma^2}=\frac{\gamma^2_k {\tilde s}^{\ast}_k {\tilde s}_k}{{\beta}^2_k\left \|{\bar{\bf V}}_{k}{\bf x}_{\rm c}\right\|_{2}^{2}+\sigma^2},
	\end{align}
	where ${\tilde s}_k\in\mathcal{D}_k$. Since ${\tilde s}^{\ast}_k {\tilde s}_k\geq s^{\ast}_k s_k=1$, the lower bound of ${{\rm SINR}}_k$ can be stated as ${{\rm SINR}}_k\geq \Gamma_k$, where 
	\begin{align}
		\Gamma_k = \frac{\gamma^2_k}{{\beta}^2_k\left \|{\bar{{\bf V}}}_{k}{\bf x}_{\rm c}\right\|_{2}^{2}+\sigma^2}.
	\end{align}
	{Since this SINR lower bound is dependent on $\gamma_k$ and ${\bf x}_{\rm c}$, it can be improved by optimizing these two variables.} Therefore, the formulation of the SINR balancing problem can be reformulated as 
	\begin{align}
		\begin{split}
			&\max \limits _{{\bf x}_{\rm c},\gamma_1,\cdots,\gamma_K} \min_{k \in {\mathcal{ K}}} \,\,\, \Gamma_k\\  \text{s.t.}& \;\; {\bar{\bf h}}^T_k{{\bf x}_{\rm c}}\in \gamma_k\cdot\mathcal{D}_k,\;\forall k \in {\mathcal{ K}}, \\
			&\;\;   \left \| {\bf x}_{\rm c}\right\|_{2}^{2} \leq P_{\rm T},
		\end{split}
		\label{robust SB}
	\end{align}
	where $P_{\rm T}$ is the transmit power budget.
	
	\begin{remark}
	For the quasi-static scenario, we have $\alpha_k\to 1$ that leads to ${\beta}^2_k \|{\bar{{\bf V}}}_{k}{\bf x}_{\rm c}\|_{2}^{2} \to 0$ and ${{\bar{\bf h}}}_k\to {{\bf h}}_k$, based on which problem \eqref{robust SB} of SINR balancing degenerates into the problem of conventional SINR balancing with perfect CSI \cite{7103338,8466792,Li2021,7042789}:
		\begin{align} 
			&\max \limits _{{\bf x}_{\rm c},\gamma} \,\,\, \gamma  \nonumber\\ \quad \text{s.t.}& \;\; {{\bf h}}_k^T{{\bf x}_{\rm c}}\in \gamma\cdot\mathcal{D}_k,\;\forall k \in {\mathcal{ K}}, \nonumber\\
			&\;\;   \left \| {\bf x}_{\rm c}\right\|_{2}^{2} \leq P_{\rm T}.
			\label{conventional-SB2}
		\end{align}
		\label{CISB remark}
	\end{remark}

	For the convenience of real representation, we define 
	\begin{align}
		\begin{split}
			&{{\bf x}}  =
			\begin{bmatrix}
				\real({\bf x}_{\rm c})\\
				\imaginary({\bf x}_{\rm c})
			\end{bmatrix},
			{{\bf s}} = 
			\begin{bmatrix}
				\real({\bf s}_{\rm c})\\
				\imaginary({\bf s}_{\rm c})
			\end{bmatrix}, 
			{{\bf V}}_k = \begin{bmatrix}
				\real({\bar{{\bf V}}}_{k})  & -\imaginary({\bar{{\bf V}}}_{k})\\
				\imaginary({\bar{{\bf V}}}_{k})  & \real({\bar{{\bf V}}}_{k})
			\end{bmatrix},\\
			&\quad
			{\bar{\bf H}} = 
			\begin{bmatrix}
				{\bar{\bf h}}^T_1 &
				\cdots &
				{\bar{\bf h}}^T_K
			\end{bmatrix}^T,
			{{\bf H}} = 
			\begin{bmatrix}
				\real({\bar{\bf H}}) & -\imaginary({\bar{\bf H}})\\
				\imaginary({\bar{\bf H}}) & \real({\bar{\bf H}})
			\end{bmatrix},
		\end{split}
	\end{align}
	and it is easy to verify that 
	\begin{align}
		\left \|{\bf V}_{k}{{\bf x}}\right\|_{2}^{2} = \left \|{\bar{{\bf V}}}_{k}{\bf x}_{\rm c}\right\|_{2}^{2},\ \left \|{{\bf x}}\right\|_{2}^{2}=\left \|{\bf x}_{\rm c}\right\|_{2}^{2}.
		\label{norm}
	\end{align}
	According to the CIR description in \cite{9910472}, the problem of SINR balancing \eqref{robust SB} can be reformulated as 
	\begin{align}
		\begin{split}
			&\max \limits_{{{\bf x}},{\boldsymbol{\delta}},\gamma_1,\cdots,\gamma_K} \min_{k \in {\mathcal{ K}}} \,\,\, \frac{\gamma^2_k}{{\beta}^2_k\left\|{\bf V}_{k}{{\bf x}}\right\|_{2}^{2}+\sigma^2}\\  
			&\quad\text{s.t.} \;\; {{\bf H}}{{\bf x}}={{\boldsymbol{\Gamma}}}\left({{\bf s}}+{\boldsymbol{\Lambda}}{\boldsymbol{\delta}}\right),\\
			&\quad\quad\ \ \left\|{{\bf x}}\right\|_{2}^{2} \leq P_{\rm T},\\
			&\quad\quad\ \ {\boldsymbol{\delta}}\succeq {\bf 0},\\
			&\quad\quad\ \ \gamma_k>0,  \;\forall k \in {\mathcal{ K}},
		\end{split}
		\label{SINR lower bound 2}
	\end{align}
	where ${{\boldsymbol{\Gamma}}}={\bf I}_2\otimes{\rm diag}\{\gamma_1, ..., \gamma_K\}$ and {${\boldsymbol{\Lambda}}$ is a matrix composed by diagonal matrices with the real and imaginary parts of $\mu_1,...,\mu_K,\nu_1,...,\nu_K$ as their elements}. $\gamma_k$ is constrained to be positive since the PSK demodulator may directly demodulate the received signal by judging its phase without rescaling. 
	
	{
	\begin{ppn}
		The first and third constraints in problem \eqref{SINR lower bound 2} can be simplified to the following constraint
		\begin{align}
			{\rm s.t.}\ {\boldsymbol{\Lambda}}^{-1}\left({{\bf H}}{{\bf x}}-{{\boldsymbol{\Gamma}}}{{\bf s}}\right)\succeq {\bf 0}.
		\end{align}\label{constraint transformation}
		\end{ppn}
		\begin{pf}
		See Appendix \ref{constraint transformation proof}.
		\end{pf}
	}

	{The above proposition removes the variable ${\boldsymbol{\delta}}$ and thus reduces the dimensionality of the variables to be optimized.} Furthermore, we take the square root of the objective function, and the above problem can be further expressed as 
		\begin{align}
			\begin{split}
				&\max \limits_{{\bf x},\gamma_1,\cdots,\gamma_K} \min_{k \in {\mathcal{ K}}} \,\,\, \frac{\gamma_k}{\sqrt{{\beta}^2_k{{\bf x}}^T{\bf E}_k{{\bf x}}+\sigma^2}}\\  
				&\quad\text{s.t.} \;\; {\boldsymbol{\Lambda}}^{-1}\left({{\bf H}}{{\bf x}}-{{\boldsymbol{\Gamma}}}{{\bf s}}\right)\succeq {\bf 0},\\
				&\quad\quad\ \ \left\| {{\bf x}}\right\|_{2}^{2} \leq P_{\rm T},
			\end{split}
			\label{robust SINR balancing problem}
		\end{align}
		where ${\bf E}_k={\bf V}_{k}^T{\bf V}_{k}$. $\gamma_k$ and $\sqrt{{\beta}^2_k{{\bf x}}^T{\bf E}_k{{\bf x}}+\sigma^2}$ in the objective function are concave and convex functions for variables, respectively. Although we consider $M$-PSK ($M>2$), the problem for $M=2$ can also be transformed into a similar form, where ${\boldsymbol{\Lambda}}^{-1}$ is replaced by $[{\rm diag}\{\real({\bf s}_{\rm c})\}\ {\rm diag}\{\imaginary({\bf s}_{\rm c})\}]$. Problem \eqref{robust SINR balancing problem} can be considered a typical MMFP problem with a quasi-concave objective and can be solved using the GD algorithm \cite{zappone2015energy}.
	The basic idea of the GD algorithm lies in obtaining the global solution by solving a sequence of the following problems:
	\begin{align}
		\max\limits_{{{\bf x}},\gamma_1,\cdots,\gamma_K\in \mathcal{S}}\left\{ \min\limits_{k \in {\mathcal{ K}}} \left\{\gamma_k-\lambda \sqrt{\beta^2_k{{\bf x}}^T{\bf E}_k{{\bf x}}+\sigma^2}  \right\} \right\},
		\label{auxiliary function}
	\end{align}
	where $\lambda\in{\mathbb{R}}$ is an intermediate variable, and $\mathcal{S}$ denotes the feasible set defined by the constraints in \eqref{robust SINR balancing problem}.
	The above problem is a convex max-min fairness problem and can be solved with generic convex optimization algorithms \cite{9687565}, e.g., the interior point algorithm \cite{byrd2000trust}, whose complexity order is $\mathcal{O}\left((N+K)^3\right)$ per iteration. Although the computation complexity of ${\bf E}_k$ scales as $\mathcal{O}\left(F_{\rm vh}N^3\right)$, it is required to be calculated only once during the long period when ${\bf m}_k$ remains unchanged \cite{9516008}.

	\subsection{Low-Complexity Design for SINR Balancing-Based SLP}\label{LC robust SINR}
	In the scenario of massive MIMO, the number of transmit antennas $N$ becomes very large, leading to a rapid rise in computational complexity for the SLP scheme developed in the previous subsection. To address this issue, we make an approximation of ${\bf E}_k$ to simplify the SINR balancing problem and design a low-complexity SLP scheme in this subsection.
	
	Due to ${\bf E}_k={\bf V}_{k}^T{\bf V}_{k}$, we have
	\begin{align}
		{\bf E}_k = 
		\begin{bmatrix}
			\real({{\bar{\bf V}}_{k}}^H{{\bar{\bf V}}_{k}}) & -\imaginary({{\bar{\bf V}}_{k}}^H{{\bar{\bf V}}_{k}})\\
			\imaginary({{\bar{\bf V}}_{k}}^H{{\bar{\bf V}}_{k}}) & \real({{\bar{\bf V}}_{k}}^H{{\bar{\bf V}}_{k}})
		\end{bmatrix},
	\end{align}
	where
	\begin{align}
		[{{\bar{\bf V}}_{k}}^H{{\bar{\bf V}}_{k}}]_{i,j} = \sum_{n=1}^{F_{\rm vh}N}[{\bf m}_k]^{2}_{n} [{\bf v}_i]^{\ast}_{n}[{\bf v}_j]_{n}.
	\end{align}

	Based on the standard orthogonality of $\{{\bf v}_n\}^N_{n=1}$, we have the following properties of ${\bf E}_k$:
	\begin{itemize}
		\item $[{\bf E}_k]_{n,n}\geq |[{\bf E}_k]_{n,j}|$,  $[{\bf E}_k]_{n,n}\geq |[{\bf E}_k]_{j,n}|$, $\forall i,j \neq n$.
		\item $[{\bf E}_k]_{i,j}\to 0$, $\forall i\neq j$, when ${\bf m}_k\to a{\bf 1},a\in{\mathbb{R}}^{+}$.
		\item ${\rm tr}({\bf E}_k)=2\|{\bf m}_k\|^2_2$.
	\end{itemize}	
	These properties indicate that the diagonal elements of ${\bf E}_k$ {are dominant in the sense that they have the highest energy in each row/column}, and as the element values of ${\bf m}_k$ become similar, ${\bf E}_k$ approaches a diagonal matrix. Thus, for simplification, we make the approximation ${\hat{\bf E}}_k= (\|{\bf m}_k\|^2_2/{N})\cdot{\bf I}_{2N}$. Furthermore, as the power of the optimal ${\bf x}$ in \eqref{SINR lower bound 2} can be proven to be $P_{\rm T}$, we constrain $\left\|{{\bf x}}\right\|^2_2 = P_{\rm T}$ to simplify the objective function, based on which the approximation of SINR balancing \eqref{SINR lower bound 2} can be represented by
	\begin{align}
		\begin{split}
			&\max \limits_{{{\bf x}},{\boldsymbol{\delta}},\gamma_1,\cdots,\gamma_K} \min_{k \in {\mathcal{ K}}} \,\,\, \frac{\gamma^2_k}{{\beta}^2_k\|{\bf m}_k\|^2_2P_{\rm T}/N+\sigma^2}\\  
			&\quad\text{s.t.} \;\; {{\bf H}}{{\bf x}}={{\boldsymbol{\Gamma}}}\left({{\bf s}}+{\boldsymbol{\Lambda}}{\boldsymbol{\delta}}\right),\\
			&\quad\quad\ \ \left\|{{\bf x}}\right\|_{2}^{2} = P_{\rm T},\\
			&\quad\quad\ \ {\boldsymbol{\delta}}\succeq {\bf 0},\\
			&\quad\quad\ \ \gamma_k>0,  \;\forall k \in {\mathcal{ K}}.
		\end{split}
		\label{robust SINR balancing LC1}
	\end{align}
	Similar to \eqref{robust SINR balancing problem}, the above problem has a quasi-concave objective with more than one optimal solution.
	
	\begin{ppn}\label{closed form T}
	The following solution of ${\boldsymbol{\delta}}^{\star},\gamma^{\star}_1...,\gamma^{\star}_K,{{\bf x}}^{\star}$
	achieves the optimum of problem \eqref{robust SINR balancing LC1}, where 
	\begin{align}
	{\boldsymbol{\delta}}^{\star}&=\arg\min\limits_{{\boldsymbol{\delta}}\succeq {\bf 0}}\|\sqrt{N}{{\bf H}}^{\dagger}{{\boldsymbol{\Theta}}}\left({{\bf s}}+{\boldsymbol{\Lambda}}{\boldsymbol{\delta}}\right)\|^2_2,\label{NNLS robust SB}\\
			\gamma^{\star}_k &= \sqrt{\frac{P_{\rm T}}{\|{{\bf H}}^{\dagger}{{\boldsymbol{\Theta}}}\left({{\bf s}}+{\boldsymbol{\Lambda}}{\boldsymbol{\delta}}^{\star}\right)\|^2_2}}\cdot\tau_k, \;\forall k \in {\mathcal{ K}},\label{low-complexity SB1}\\
			{{\bf x}}^{\star} &= \sqrt{\frac{P_{\rm T}}{\|{{\bf H}}^{\dagger}{{\boldsymbol{\Theta}}}\left({{\bf s}}+{\boldsymbol{\Lambda}}{\boldsymbol{\delta}}^{\star}\right)\|^2_2}}{{\bf H}}^{\dagger}{{\boldsymbol{\Theta}}}\left({{\bf s}}+{\boldsymbol{\Lambda}}{\boldsymbol{\delta}}^{\star}\right),\label{low-complexity SB2}
	\end{align}
	where ${{\bf H}}^{\dagger}={{\bf H}}^T({{\bf H}}{{\bf H}}^T)^{-1}$, 
	$\tau_k=\sqrt{{\beta}^2_k\|{\bf m}_k\|^2_2P_{\rm T}/N+\sigma^2}$, and ${{\boldsymbol{\Theta}}} = {\bf I}_{2}\otimes{\rm diag}\{\tau_1,...,\tau_K\}$.
	\end{ppn}
	\begin{pf}
	See Appendix \ref{closed form T proof}
	\end{pf}
	
	\begin{remark}
			\ppnref{closed form T} simplifies the process of solving problem \eqref{robust SINR balancing LC1} by reducing it to solving problem \eqref{NNLS robust SB} and computing the closed-form expressions given by \eqref{low-complexity SB1}-\eqref{low-complexity SB2}, where problem \eqref{NNLS robust SB} is a non-negative least squares (NNLS) problem that can be efficiently solved \cite{lawson1995solving}. This also implies that the key to solving \eqref{robust SINR balancing LC1} is to find the optimal ${\boldsymbol{\delta}}^{\star}$.
		\end{remark}

	Moreover, we analyze the properties of problem \eqref{NNLS robust SB} and its optimal solution ${\boldsymbol{\delta}}^{\star}$ in a special case, providing ideas for further reducing computational complexity.
	
	\begin{lemma}\label{L1}
	${\boldsymbol{\delta}}^{\star}={\bf 0}$ achieves the optimum of the NNLS problem of the following form if ${\bf R}_{\bf A}={\bf A}^T{\bf A}$ is a diagonal matrix.
		\begin{align}
			\min\limits_{{\boldsymbol{\delta}}\succeq {\bf 0}}\|{\bf A}\left({{\bf s}}+{\boldsymbol{\Lambda}}{\boldsymbol{\delta}}\right)\|^2_2.
			\label{a NNLS form}
		\end{align}
	\end{lemma}
	\begin{pf}
	See Appendix \ref{Lemma 1 proof}.
	\end{pf}
	
	\begin{ppn}\label{SB P}
	For NNLS problem \eqref{NNLS robust SB}, ${\bf R}_{\bf A}$ in form \eqref{a NNLS form} converges into the diagonal matrix $({\bf I}_2\otimes{\bf M}^{-1}){{\boldsymbol{\Theta}}}^2$ in probability when $N\gg K$ with the assumptions in \cite{8815518}, where
	\begin{align}
		{\bf M}={\rm diag}\left\{\frac{{\alpha}^2_{1}\|{\bf m}_{1}\|^2_2}{N},...,\frac{{\alpha}^2_{K}\|{\bf m}_{K}\|^2_2}{N}\right\}.
	\end{align}
	\end{ppn}
	\begin{pf}
	See Appendix \ref{SB NNLS P proof}.
	\end{pf}
	
	From \lmref{L1} and \ppnref{SB P}, it can be concluded that the optimal ${\boldsymbol{\delta}}^{\star}$ converges to ${\bf 0}$ in most practical scenarios when $N\gg K$.
	To further reduce the computational complexity, we approximate
	$\gamma^{\star}_k$ and ${{\bf x}}^{\star}$ as the closed forms \eqref{low-complexity SB1} and \eqref{low-complexity SB2}, where ${\boldsymbol{\delta}}^{\star}$ is approximated by ${\bf 0}$. 
	
	The low-complexity SLP scheme developed in this subsection is summarized as \algref{closed form SB}, whose computational complexity mainly depends on steps 4 and 5 with the order of $\mathcal{O}\left(K^2(N+K)\right)$. {It is worth noting that, the low-complexity algorithm approximates ${\bf E}_k$ as ${\hat {\bf E}}_k$ and ignores the interference with average power ${\beta}^2_k{{\bf x}}^T({\bf E}_k-{\hat {\bf E}}_k){{\bf x}}$, leading to a decrease in performance compared to the original one. By contrasting it with the expression of the total interference plus noise power ${\beta}^2_k{{\bf x}}^T{\bf E}_k{{\bf x}}+\sigma^2$, it can be concluded that, for a given ${\bf x}$, the performance gap caused by the ignored interference gradually widens as ${\beta}^2_k$ increases or $\sigma^2$ decreases.}
	
	\begin{algorithm}[t]
			% \setstretch{1.35}
			\caption{Closed-Form Robust SLP Based on SINR Balancing Criterion}
			\label{closed form SB}
			\begin{spacing}{1.3}
				\begin{algorithmic}[1]
					\STATE \textbf{Input:} ${{\bf H}}$, $\{{\bf m}_k, {\beta}_{k}\}^K_{k=1}$, ${{\bf s}}$, $\sigma^2$, $P_{\rm T}$.
					\STATE $\tau_k = \sqrt{{\beta}^2_k\|{\bf m}_k\|^2_2P_{\rm T}/N+\sigma^2}, \ \forall k\in{\mathcal{ K}}$.
					\STATE ${{\boldsymbol{\Theta}}} = {\bf I}_{2}\otimes{\rm diag}\{\tau_1,...,\tau_K\}$.
					\STATE ${{\bf H}}^{\dagger}={{\bf H}}^T({{\bf H}}{{\bf H}}^T)^{-1}$.
					\STATE
					${{\bf x}} = \sqrt{\frac{P_{\rm T}}{\|{{\bf H}}^{\dagger}{{\boldsymbol{\Theta}}}{{\bf s}}\|^2_2}}\cdot{{\bf H}}^{\dagger}{{\boldsymbol{\Theta}}}{{\bf s}}$.
					\STATE $\gamma_k = \sqrt{\frac{P_{\rm T}}{\|{{\bf H}}^{\dagger}{{\boldsymbol{\Theta}}}{{\bf s}}\|^2_2}}\cdot\tau_k, \ \forall k\in{\mathcal{ K}}$.
					\STATE \textbf{Output:}  ${{\bf x}},\gamma_1, ..., \gamma_K$.
				\end{algorithmic}
			\end{spacing}
		\end{algorithm}

	\section{Robust Symbol-Level Precoding With MMSE Criterion}\label{robust MMSE section}
	{{Apart from considering the SINR balancing problem, which tends to improve the transmission quality (SINR lower bound) for the worst user, we further investigate the MMSE problem in this section, which aims to enhance the average performance of all users, i.e., minimizes the MSE between the signal to be demodulated and the expected signal \cite{9910472}.}}
	\subsection{Problem Statement}
	We firstly formulate the MMSE problem for robust SLP transmission, which minimizes the MSE between the signal to be demodulated $({\bar{\bf h}}_{k}^{T}{\bf x}_{\rm c} + {\bar{n}}_k)/\gamma_k$ and the expected signal ${\tilde{s}}_k$ located in $\mathcal{D}_k$, i.e., 
	\begin{align}
		\begin{split}
			&\min_{{\bf x}_{\rm c}, {\tilde{\bf s}}_{\rm c},\gamma_1,...,,\gamma_K}\  {\mathbb{E}}_{{{\bf n}}}\left\{\left\|{\bar{\boldsymbol{\Gamma}}}^{-1} \left({{\bar{\bf H}}{\bf x}_{\rm c}+{\bar{\bf n}}}\right)-{\tilde{\bf s}}_{\rm c}\right\|^2_2\right\}\\
			&\quad{\rm s.t.}\ \left\|{\bf x}_{\rm c}\right\|^2_2 \leq P_{\rm T},\\
			&\quad\quad\quad\tilde{ s}_k\in\mathcal{D}_k,\ \forall k \in {\mathcal{ K}},\\
			&\quad\quad\quad\gamma_k>0,  \;\forall k \in {\mathcal{ K}},
		\end{split}
		\label{robust MMSE SLP 2}
	\end{align}
	where ${\bar{\bf n}}=[{\bar n}_1, ..., {\bar n}_K]$, ${\tilde{\bf s}}_{\rm c}=[{\tilde s}_1,...,{\tilde s}_K]$, and ${\bar{\boldsymbol{\Gamma}}}={\rm diag}\{\gamma_1, ..., \gamma_K\}$.
	
	\begin{remark}
	For the quasi-static scenario, by forcing $\gamma_1=\cdots=\gamma_K=\gamma$, the MMSE problem \eqref{robust MMSE SLP 2} for robust SLP will degenerate into the conventional MMSE problem with perfect CSI \cite{9910472}:
		\begin{align}
			\begin{split}
				&\min\limits_{{\bf x}_{\rm c}, {\tilde{\bf s}}_{\rm c}, \gamma} \ \mathbb{E}_{{\bf n}_{\rm c}}\left\{\left \|\frac{{\bf H}_{\rm c}{\bf x}_{\rm c}+{\bf n}_{\rm c}}{\gamma}-{\tilde{\bf s}}_{\rm c}\right\|_2^2\right\}	
				\\
				&\quad{\rm s.t.}\ \left\|{\bf x}_{\rm c}\right\|^2_2 \leq P_{\rm T},\\
				&\quad\quad\quad {\tilde{s}}_k\in \mathcal{D}_k,\;\forall k\in{\mathcal{ K}},
			\end{split}                                                        
			\label{CI-MMSE}
		\end{align}	
		where  ${\bf H}_{\rm c} = \left[{\bf h}_1^{T}, ..., {\bf h}_K^{T}\right]^T$ and ${\bf n}_{\rm c} = \left[n_{1}, ..., n_{K}\right]^T$. {Given a fixed transmit power, the optimal solution to \eqref{CI-MMSE} will become the optimal solution to \eqref{conventional-SB2} as $\sigma^2$ approaches zero \cite{9910472}. This implies that the SLP schemes based on these two conventional problems will gradually approach each other at extremely high signal-to-noise ratio (SNR).}
		\label{CIMMSE remark}
	\end{remark}
	
	Although the MMSE problem \eqref{robust MMSE SLP 2} is or can be transformed into a convex problem for one of ${\bf x}_{\rm c}$, ${\tilde{\bf s}}_{\rm c}$, and $\gamma_k$, it is not convex or quasiconvex when optimizing these variables jointly. Consequently, we derive the optimal solution forms of its subproblems separately and propose an alternating optimization algorithm in the following subsection.
	\subsection{Alternating Optimization Algorithm}
	\label{CIMMSER-Algorithm}
	\begin{figure*}[hb]
		\normalsize
		\setcounter{MYtempeqncnt}{\value{equation}}
		\vspace*{4pt}
		\hrulefill
		\begin{equation}
			\begin{split}
				f_{\rm p1} &= {\tilde{\bf s}}^T{\bf P}^T{{\bf H}}^T{{\boldsymbol{\Psi}}}^2{{\bf H}}{\bf P}{\tilde{\bf s}}-2{\tilde{\bf s}}{{\boldsymbol{\Psi}}}{{\bf H}}{\bf P}{\tilde{\bf s}}+{\tilde{\bf s}}^T{\tilde{\bf s}}+\frac{{\sigma^2}\sum_{k=1}^{K}{\psi^2_k}}{P_{\rm T}}{\tilde{\bf s}}^T{\bf P}^T{\bf P}{\tilde{\bf s}}+{\tilde{\bf s}}^T{\bf P}^T{\boldsymbol {\Upsilon}}{\bf P}{\tilde{\bf s}}\\
				& = {\tilde{\bf s}}^T\left({\bf P}^T{{\bf H}}^T{{\boldsymbol{\Psi}}}^2{{\bf H}}{\bf P}-2{{\boldsymbol{\Psi}}}{{\bf H}}{\bf P}+{\bf I}_{2K}+\frac{{\sigma^2}\sum_{k=1}^{K}{\psi^2_k}}{P_{\rm T}}{\bf P}^T{\bf P}+{\bf P}^T{\boldsymbol {\Upsilon}}{\bf P}\right){\tilde{\bf s}}\\
				& = {\tilde{\bf s}}^T\left[{\bf P}^T\left({{\bf H}}^T{{\boldsymbol{\Psi}}}^2{{\bf H}}+\frac{{\sigma^2}\sum_{k=1}^{K}{\psi^2_k}}{P_{\rm T}}{\bf I}_{2N}+{\boldsymbol {\Upsilon}}\right){\bf P}-2{{\boldsymbol{\Psi}}}{{\bf H}}{\bf P}+{\bf I}_{2K}\right]{\tilde{\bf s}}.
			\end{split}
			\tag{\ref{robust MMSE SLP sub1 f2}-A}
			\label{robust MMSE SLP sub1 f1}
		\end{equation}
		% Restore the current equation number.
		\setcounter{equation}{\value{MYtempeqncnt}}
	\end{figure*}
	
	Define the variables ${\psi_k},\ \forall k \in {\mathcal{ K}}$ and $\eta$, which satisfy 
	\begin{align}
		\frac{\psi_k}{\eta}=\frac{1}{\gamma_k},\ \forall k \in {\mathcal{ K}},
	\end{align}
	and the MMSE problem \eqref{robust MMSE SLP 2} can be reformulated as
	\begin{align}
		\begin{split}
			&\min_{{\bf x}_{\rm c}, {\tilde{\bf s}}_{\rm c},{\psi_1},...,,{\psi_K}, \eta}\  {\mathbb{E}}_{{\bar{\bf n}}}\left\{\left\|\frac{\bar{\boldsymbol{\Psi}} \left({{\bar{\bf H}}{\bf x}_{\rm c}+{\bar{\bf n}}}\right)}{\eta}-{\tilde{\bf s}}_{\rm c}\right\|^2_2\right\}\\
			&\quad{\rm s.t.}\ \left\|{\bf x}_{\rm c}\right\|^2_2 \leq P_{\rm T},\\
			&\quad\quad\quad\tilde{s}_k\in\mathcal{D}_k,  \;\forall k \in {\mathcal{ K}},\\
			&\quad\quad\quad\frac{\psi_k}{\eta}>0,  \;\forall k \in {\mathcal{ K}},
		\end{split}
		\label{robust MMSE SLP 3}
	\end{align}
	where ${\bar{\boldsymbol{\Psi}}}={\rm diag}\{\psi_1, ..., \psi_K\}$. By introducing the CIR description in \cite{9910472} and defining 
	\begin{align}
	{{\boldsymbol{\Psi}}}=
		{\begin{bmatrix}
				{\bar{\boldsymbol{\Psi}}} &  \\
				& {\bar{\boldsymbol{\Psi}}}
		\end{bmatrix}}, 
		{{\bf n}} = 
		\begin{bmatrix}
			\real({\bar{\bf n}})\\
			\imaginary({\bar{\bf n}})
		\end{bmatrix},
	\end{align}
	the real representation of problem \eqref{robust MMSE SLP 3} is given by 
	\begin{align}
		\begin{split}
			&\min_{{{\bf x}}, {\boldsymbol{\delta}},{\psi_1},...,,{\psi_K}, \eta}\  {\mathbb{E}}_{{{\bf n}}}\left\{\left\|\frac{{{\boldsymbol{\Psi}}} \left({{{\bf H}}{{\bf x}}+{{\bf n}}}\right)}{\eta}-\left({{\bf s}}+{\boldsymbol{\Lambda}}{\boldsymbol{\delta}}\right)\right\|^2_2\right\}\\
			&\quad{\rm s.t.}\ \left\|{{\bf x}}\right\|^2_2 \leq P_{\rm T},\\
			&\quad\quad\quad\frac{\psi_k}{\eta}>0,  \;\forall k \in {\mathcal{ K}},\\
			&\quad\quad\quad{\boldsymbol{\delta}}\succeq {\bf 0}.
		\end{split}
		\label{robust MMSE SLP 4}
	\end{align}
	Motivated by \cite{4712693} and \cite{7397861}, we try to analyze the property of this problem by fixing some variables and finding the optimal solution for the remaining variables. 
	
	With fixed ${\psi_1},...,,{\psi_K}>0$, problem \eqref{robust MMSE SLP 4} is transformed into the following problem:
	\begin{align}
		\begin{split}
			&\min_{{{\bf x}}, {\boldsymbol{\delta}}, \eta}\  {\mathbb{E}}_{{{\bf n}}}\left\{\left\|\frac{{{\boldsymbol{\Psi}}} \left({{{\bf H}}{{\bf x}}+{{\bf n}}}\right)}{\eta}-\left({{\bf s}}+{\boldsymbol{\Lambda}}{\boldsymbol{\delta}}\right)\right\|^2_2\right\}\\
			&\quad{\rm s.t.}\ \left\|{{\bf x}}\right\|^2_2 \leq P_{\rm T},\\
			&\quad\quad\quad{\boldsymbol{\delta}}\succeq {\bf 0},\ \eta>0.
		\end{split}
		\label{robust MMSE SLP sub1}
	\end{align}	
	
	\begin{ppn}\label{CIMMSE P}
	For certain ${\boldsymbol{\delta}}$, the optimal ${{\bf x}}^{\star}, \eta^{\star}$ in problem \eqref{robust MMSE SLP sub1} can be expressed as
		\begin{align}
			\begin{split}
				&{{\bf x}}^{\star}=\eta^{\star}{\bf P}{\tilde{\bf s}},\\
				&\eta^{\star} = \sqrt{\frac{P_{\rm T}}{\left\|{\bf P}{\tilde{\bf s}}\right\|^2_2}},
			\end{split}
		\end{align}
		where
		\begin{align}
			&{\bf P} = \left({{\bf H}}^T{{\boldsymbol{\Psi}}}^2{{\bf H}}+{\boldsymbol {\Upsilon}}+\frac{{\sigma^2}\sum_{k=1}^{K}{\psi^2_k}}{P_{\rm T}}{\bf I}_{2N}\right)^{-1}{{\bf H}}^T{{\boldsymbol{\Psi}}},\\
			&{\boldsymbol {\Upsilon}} = \sum^{K}_{k=1}{\psi^2_k}{\beta}^2_k{\bf E}_k,\\
			&{\tilde{\bf s}}={{\bf s}}+{\boldsymbol{\Lambda}}{\boldsymbol{\delta}}.
		\end{align}
	\end{ppn}
	\begin{pf}
	See Appendix \ref{Proposition 2 proof}.
	\end{pf}
	\begin{remark}
	Based on \ppnref{CIMMSE P}, the optimal ${{\bf x}}^{\star}, \eta^{\star}$ for a certain ${\boldsymbol{\delta}}$ can be obtained by these closed-form expressions. This greatly simplifies the process of solving problem \eqref{robust MMSE SLP sub1} since it is equivalent to finding the optimal ${\boldsymbol{\delta}}^{\star}$.
	\end{remark}
	
	Based on \ppnref{CIMMSE P}, the objective function of \eqref{robust MMSE SLP sub1} is transformed into \eqref{robust MMSE SLP sub1 f1} and \eqref{robust MMSE SLP sub1 f2}
	\begin{align}
		f_{\rm p1} = {\tilde{\bf s}}^T\left({\bf I}_{2K}-{{\boldsymbol{\Psi}}}{{\bf H}}{\bf P}\right){\tilde{\bf s}}.
		% \tag{\ref{robust MMSE SLP sub1 f2}-B}
		\label{robust MMSE SLP sub1 f2}
	\end{align}
	Then, the optimal ${\boldsymbol{\delta}}^{\star}$ in \eqref{robust MMSE SLP sub1} can be obtained by solving 
	\begin{align}
		{\boldsymbol{\delta}}^{\star} = \arg\min_{{\boldsymbol{\delta}}\succeq{\bf 0}} \|{\bf B}\left({\boldsymbol{\Lambda}}{\boldsymbol{\delta}}+{{\bf s}}\right)\|^2_2,
	\end{align}
	where ${\bf B}$ is obtained from the Cholesky decomposition
	${\bf B}^{T}{\bf B} = N({\bf I}_{2K}-{{\boldsymbol{\Psi}}}{{\bf H}}{\bf P})$. By defining ${\bf C}={\bf B}{\boldsymbol{\Lambda}}$ and ${\bf d} = -{\bf B}{{\bf s}}$, the above problem can be further expressed as 
	\begin{align}
		{\boldsymbol{\delta}}^{\star} =\arg\min_{{\boldsymbol{\delta}}\succeq{\bf 0}} \|{\bf C}{\boldsymbol{\delta}}-{\bf d}\|^2_2.
		\label{robust MMSE NNLS}
	\end{align}
	The problem is a typical NNLS problem and can be efficiently solved by the active set-based algorithm \cite{lawson1995solving}.

	With fixed ${{\bf x}}, {\boldsymbol{\delta}}$, and $\eta>0$, problem \eqref{robust MMSE SLP 4} is transformed into the following problem:
	\begin{align}
		\begin{split}
			&\min_{\psi_1,...,\psi_K}\  {\mathbb{E}}_{{{\bf n}}}\left\{\left\|\frac{{{\boldsymbol{\Psi}}} \left({{{\bf H}}{{\bf x}}+{{\bf n}}}\right)}{\eta}-{\tilde{\bf s}}\right\|^2_2\right\},\\
			&\quad{\rm s.t.}\ \psi_k>0,  \;\forall k \in {\mathcal{ K}}.\\
		\end{split}
		\label{robust MMSE SLP sub2}
	\end{align}	
	If ${{\bf x}}$ and $\eta$ are obtained by solving problem \eqref{robust MMSE SLP sub1}, the objective function $f_{\rm p2}$ of the above problem can be expressed as the first line of \eqref{robust MMSE SLP sub1 f1}, where we reiterate ${\boldsymbol {\Upsilon}} = \sum^{K}_{k=1}{\psi^2_k}{\beta}^2_k{\bf E}_k$. It is readily to verify that the objective function is convex for $\psi_1,...,\psi_K$, and the gradient with respect to $\psi_k$ is given by
	\begin{align}
		\begin{split}
			\frac{\partial f_{\rm p2}}{\partial \psi_k} =& 2\psi_k\left([{{\bf H}}{\bf u}]^2_k+[{{\bf H}}{\bf u}]^2_{K+k}\right)\\
			&- 2\left([{\tilde{\bf s}}]_k[{{\bf H}}{\bf u}]_k+[{\tilde{\bf s}}]_{K+k}[{{\bf H}}{\bf u}]_{K+k}\right)\\
			&+\frac{2\psi_k\sigma^2}{P_{\rm T}}{\bf u}^T{\bf u}+2{\psi_k}{\beta}^2_k{\bf u}^T{\bf E}_k{\bf u},
		\end{split}
	\end{align}
	where ${\bf u} = {\bf P}{\tilde{\bf s}}$. By vanishing the gradient, we can obtain the following solution:
	\begin{align}
		\psi_k = \frac{[{\tilde{\bf s}}]_k[{{\bf H}}{\bf u}]_k+[{\tilde{\bf s}}]_{K+k}[{{\bf H}}{\bf u}]_{K+k}}{\left([{{\bf H}}{\bf u}]^2_k+[{{\bf H}}{\bf u}]^2_{K+k}\right)+\frac{\sigma^2}{P_{\rm T}}{\bf u}^T{\bf u}+{\beta}^2_k{\bf u}^T{\bf E}_k{\bf u}}.
		\label{optimal psi}
	\end{align}
	The above $\psi_k$ is the optimum $\psi^{\star}_k$ if $\psi_k>0$. 
	{Therefore, we update \eqref{optimal psi} only when $\psi_k>0$ to ensure the improvement of the objective function in this iteration.}
	
	Based on the above derivations, the following alternating optimization algorithm can be easily concluded: Firstly, initialize $\psi_1,...,\psi_K$ appropriately; Secondly, solve problem \eqref{robust MMSE SLP sub1}; Third, solve problem \eqref{robust MMSE SLP sub2}; Then, iteratively perform step 2 and step 3 until some stopping criterion is met. The detailed algorithm is described in \algref{A4}. {Overall, this algorithm can iteratively optimize variables to achieve a stationary point for problem \eqref{robust MMSE SLP 4}.}
	The convergence of \algref{A4} can be easily verified as the objective function of problem \eqref{robust MMSE SLP 4} is alternately reduced until convergence. 
	
	\begin{algorithm}[htb]
		\caption{Alternating Optimization Algorithm for MMSE Problem \eqref{robust MMSE SLP 4} }
		\label{A4}
		\begin{spacing}{1.2}
			\begin{algorithmic}[1]
				\STATE \textbf{Input:} ${{\bf H}}$, $\{{{{\bf E}}}_{k},{\beta}_k\}^K_{k=1}$, ${{\bf s}}$, ${\boldsymbol{\Lambda}}$, $\sigma^2$, $P_{\rm T}$, ${\rm Iter}_{\rm max}$.
				\STATE Initialize $\psi_1=\cdots=\psi_K=1$, $n=0$.
				\STATE \textbf{while} $n<{\rm Iter}_{\rm max}$ \textbf{do}
				\STATE \quad ${{{\boldsymbol{\Psi}}}} = {\bf I}_2\otimes{\rm diag}\{\psi_1, \cdots, \psi_K\}$.
				\STATE \quad ${\boldsymbol {\Upsilon}} = \sum^{K}_{k=1}{\psi^2_k}{\beta}^2_k{\bf E}_k$.
				\STATE \quad ${\bf P} = \left({{\bf H}}^T{{{\boldsymbol{\Psi}}}}^2{{\bf H}}+{\boldsymbol {\Upsilon}}+\frac{{\sigma^2}\sum_{k=1}^{K}{\psi_k^2}}{P_{\rm T}}{\bf I}_{2N}\right)^{-1}{{\bf H}}^T{{\boldsymbol{\Psi}}}$.
				\STATE \quad Cholesky decomposition:
				${\bf B}^{T}{\bf B} = N({\bf I}_{2K}-{{\boldsymbol{\Psi}}}{{\bf H}}{\bf P})$.
				\STATE \quad ${\bf C}={\bf B}{\boldsymbol{\Lambda}}$, ${\bf d} = -{\bf B}{{\bf s}}$.
				\STATE \quad  ${\boldsymbol{\delta}}^{\star} =\arg\min_{{\boldsymbol{\delta}}\succeq{\bf 0}} \|{\bf C}{\boldsymbol{\delta}}-{\bf d}\|^2_2$.
				\STATE \quad ${\tilde{\bf s}}={{\bf s}}+{\boldsymbol{\Lambda}}{\boldsymbol{\delta}}^{\star}$, ${\bf u} = {\bf P}{\tilde{\bf s}}$. 
				\STATE \quad \textbf{for} $k\in{\mathcal{ K}}$ \textbf{do}
				\STATE \quad \quad $\psi^{'}_k = \frac{[{\tilde{\bf s}}]_k[{{\bf H}}{\bf u}]_k+[{\tilde{\bf s}}]_{K+k}[{{\bf H}}{\bf u}]_{K+k}}{\left([{{\bf H}}{\bf u}]^2_k+[{{\bf H}}{\bf u}]^2_{K+k}\right)+\frac{\sigma^2}{P_{\rm T}}{\bf u}^T{\bf u}+{\beta}^2_k{\bf u}^T{\bf E}_k{\bf u}}$.
				\STATE \quad \quad \textbf{If} $\psi^{'}_k>0$ \textbf{do}
				\STATE \quad \quad \quad $\psi_k=\psi^{'}_k$.
				\STATE \quad \quad \textbf{end} \textbf{If}
				\STATE \quad \textbf{end} \textbf{for}
				\STATE \quad $n = n+1$.	
				\STATE  \textbf{end while}
				\STATE $\eta = \sqrt{\frac{P_{\rm T}}{\left\|{\bf P}{\tilde{\bf s}}\right\|^2_2}}$, ${{\bf x}}=\eta{\bf P}{\tilde{\bf s}}$.
				\STATE \textbf{Output:}  ${{\bf x}}, {\boldsymbol{\delta}},{\psi_1},...,,{\psi_K}, \eta$.
			\end{algorithmic}
		\end{spacing}
	\end{algorithm}
	The computational complexity of \algref{A4} mainly depends on steps 6, 7, and 8. Step 6 performs the matrix multiplication and inversion with complexity order $\mathcal{O}\left(N^2(N+K)\right)$. The step 7 and step 8 perform Cholesky decomposition and matrix multiplication with complexity order $\mathcal{O}\left(K^2(N+K)\right)$. Although step 9 iteratively solves the NNLS problem by the active set-based algorithm \cite{lawson1995solving}, its computational complexity can be ignored compared with matrix multiplications since each iteration requires very low complexity, and the practical amount of iterations is quite small \cite{9910472}. Thus, the complexity order of \algref{A4} is about $\mathcal{O}\left(N^3+N^2K+NK^2\right)$ per iteration. 
	
	\subsection{Low-Complexity Design for MMSE-Based SLP}\label{CIMMSE-Robust-LC}
	In the case of massive MIMO, i.e., $N\gg K$, we consider the low-complexity SLP based on the MMSE criterion. The computational complexity of \algref{A4} mainly depends on step 6. 
	According to Section \ref{LC robust SINR}, we have the approximation ${\hat{\bf E}}_k= (\|{\bf m}_k\|^2_2/{N})\cdot{\bf I}_{2N}$. Then, ${\boldsymbol {\Upsilon}}$ is approximated by the diagonal matrix ${\hat{\boldsymbol {\Upsilon}}} = (\sum^{K}_{k=1}{\psi^2_k}{\beta}^2_k\|{\bf m}_k\|^2_2/{N})\cdot{\bf I}_{2N}$, and the approximation of ${\bf P}$ can be expressed as
		\begin{align}
			{\hat{\bf P}}&= \left({{\bf H}}^T{{{\boldsymbol{\Psi}}}}^2{{\bf H}}+{\hat{\boldsymbol {\Upsilon}}}+\frac{{\sigma^2}\sum_{k=1}^{K}{\psi_k^2}}{P_{\rm T}}{\bf I}_{2N}\right)^{-1}{{\bf H}}^T{{\boldsymbol{\Psi}}}\\
			&=\left({{\bf H}}^T{{{\boldsymbol{\Psi}}}}^2{{\bf H}}+\kappa{\bf I}_{2N}\right)^{-1}{{\bf H}}^T{{\boldsymbol{\Psi}}},
		\end{align}
			where 
		\begin{align}
		\kappa={{\sum^{K}_{k=1}{\psi^2_k}\left(\frac{{\beta}^2_k\|{\bf m}_k\|^2_2}{N}+\frac{\sigma^2}{P_{\rm T}}\right)}}.
		\end{align}
		Based on the matrix inversion lemma (MIL), ${\hat{\bf P}}$ can be rewritten as
		\begin{align}
			{\hat{\bf P}} ={{\bf H}}^T{{\boldsymbol{\Psi}}}\left({{\boldsymbol{\Psi}}}{{\bf H}}{{\bf H}}^T{{\boldsymbol{\Psi}}}+{\kappa}{\bf I}_{2K}\right)^{-1}.
			\label{approx P}
		\end{align}
	By replacing ${\bf P}$ in \algref{A4} with ${\hat{\bf P}}$ in \eqref{approx P}, the computational complexity of step 6 is reduced to $\mathcal{O}\left(K^2(N+K)\right)$.
	
	\begin{ppn}\label{CIMMSE NNLS P}
	For NNLS problem \eqref{robust MMSE NNLS} with certain ${{\boldsymbol{\Psi}}}$ and approximation matrix ${\hat{\boldsymbol {\Upsilon}}}$, the ${\bf R}_{\bf A}$ of form \eqref{a NNLS form} converges into the diagonal matrix $\kappa\left[{{\boldsymbol{\Psi}}}^2({\bf I}_2\otimes{\bf M})\right]^{-1}$ in probability when $N\gg K$ with the assumptions in \cite{8815518}.
	\end{ppn}
	\begin{pf}
	See Appendix \ref{CIMMSE NNLS P proof}.
	\end{pf}
	
	Similar to \algref{closed form SB}, ${\boldsymbol{\delta}}^{\star}$ in problem \eqref{robust MMSE NNLS} is approximated by ${\bf 0}$ according to \textbf{{Lemma \ref{L1}}} and \ppnref{CIMMSE NNLS P} when $N\gg K$, and steps 7-9 in \algref{A4} can be omitted. In conclusion, the low-complexity SLP can be implemented by the modified \algref{A4}, whose ${\bf P}$ in step 6 is replaced by ${\hat{\bf P}}$ in \eqref{approx P}, and steps 7-9 are omitted with ${\boldsymbol{\delta}}^{\star}={\bf 0}$.
	
	Based on \lmref{L1}, \ppnref{SB P}, and \ppnref{CIMMSE NNLS P}, it can be concluded that when inter-user channels are intended to be orthogonal (e.g., $N\gg K$) in most practical scenarios, ${\boldsymbol{\delta}}^{\star}={\bf 0}$ exists for the low-complexity SLP schemes presented in Section \ref{LC robust SINR} and Section \ref{CIMMSE-Robust-LC}. Similar results were also observed in simulations of SLP schemes presented in Section \ref{Section 4} and Section \ref{CIMMSER-Algorithm}. Since ${\boldsymbol{\delta}}^{\star}={\bf 0}$ results in ${\bf{\tilde {s}}}={\bf{{s}}}$, such a phenomenon implies that the proposed schemes will not achieve a performance gain from the interference exploitation mechanism in scenarios with less interference between users. In these cases, the symbol-level optimization of rescaling factors and transmit signal given time-varying channels will become the primary source of performance improvement.
	It is also noteworthy that the conclusions and the proofs of \ppnref{SB P} and \ppnref{CIMMSE NNLS P} can be easily extended to the NNLS problems from \eqref{conventional-SB2} and \eqref{CI-MMSE} with perfect CSI, i.e., the SLP problems of conventional SINR balancing and MMSE \cite{8815429,9910472}.
	
	\section{Numerical Results}\label{Section 6}
	In this section, we employ the Monte Carlo method to assess the performance of the proposed schemes. We consider two antenna configurations: i) The BS is equipped with ULA that $N=14$ while $F_{\rm vh}=1$ and $K=12$. ii) The BS is equipped with UPA comprising $N_{\rm v}=4$ dual-polarized antennas in each column and $N_{\rm h}=8$ dual-polarized antennas in each row with the number of antennas $N=2N_{\rm v}N_{\rm h}$ while $F_{\rm vh}=4$ and $K=9$. Using a method similar to that in \cite{8694866}, $\{{\bf m}_k\}_{k=1}^K$ is generated from the channels created based on the Quadriga channel model \cite{6758357,1614066}.
	Specifically, the antennas of BS and UE are `3gpp-3d' and `omni', the center frequency is set at 3.5 GHz, and the scenario is `3GPP\_38.901\_UMa\_NLOS' \cite{2016}. Shadow fading and path loss are not considered. For the convenience of comparison, we consider the normalized channel satisfying $\mathbb{E}\left\{{\rm tr}\{{\bf H}{\bf H}^H\}\right\}=KN$, ${\rm SNR}=P_{\rm T}/{\sigma^2}$, and $\alpha_1=\cdots=\alpha_K=\alpha$ in the simulation.

	This section compares the following schemes:
	\begin{itemize}
		\item `\textbf{ZF}' and `\textbf{MMSE}': The ZF and MMSE precoding schemes with symbol-level power constraints \cite{Li2021}.
		\item `\textbf{CISB}' and `\textbf{CIMMSE}': The optimal solutions of  \eqref{conventional-SB2} and \eqref{CI-MMSE}.
		\item `\textbf{CISB-RNB}': The SINR balancing-based robust SLP scheme with assumption of norm-bounded error in \cite{7103338}.
		\item `\textbf{CISB-R}' and `\textbf{CIMMSE-R}': The proposed robust SLP schemes, i.e., the solutions of \eqref{robust SINR balancing problem} and \eqref{robust MMSE SLP 4}.
		\item `\textbf{CISB-RLC}' and `\textbf{CIMMSE-RLC}': The proposed low-complexity SLP schemes in Section \ref{LC robust SINR} and \ref{CIMMSE-Robust-LC}.
	\end{itemize}
	
	For better comparisons of MSE and SINR, 
	$\{{\bar{\bf h}}_{k}\}_{k=1}^K$ instead of $\{{\bf h}_{k}^{\rm{u}}\}_{k=1}^K$ is utilized for precoding by `ZF', `MMSE', `CISB', and `CIMMSE', while this almost has no impact on their performance. 
	Since the accurate error bound cannot be obtained in practical systems, we set $\sqrt{{\mathbb E}\{\|{\bf h}_{k}-{\bf h}_{k}^{\rm{u}}\|_2^2\}}$ as the norm of the $k$-th user's maximum channel error bound of `CISB-RNB' in \cite{7103338}.
	$\Gamma_{{\rm min}}=\min_k \Gamma_k$ is chosen as one of the performance evaluations, which also includes SER and the MSE between the signal to be demodulated ($y_k/\gamma_k$) and the target signal ($s_k$ or ${\tilde{s}}_k$) in SLP schemes. 
	
	\begin{figure}[htp]
		\centering
		\includegraphics[width=3.5in]{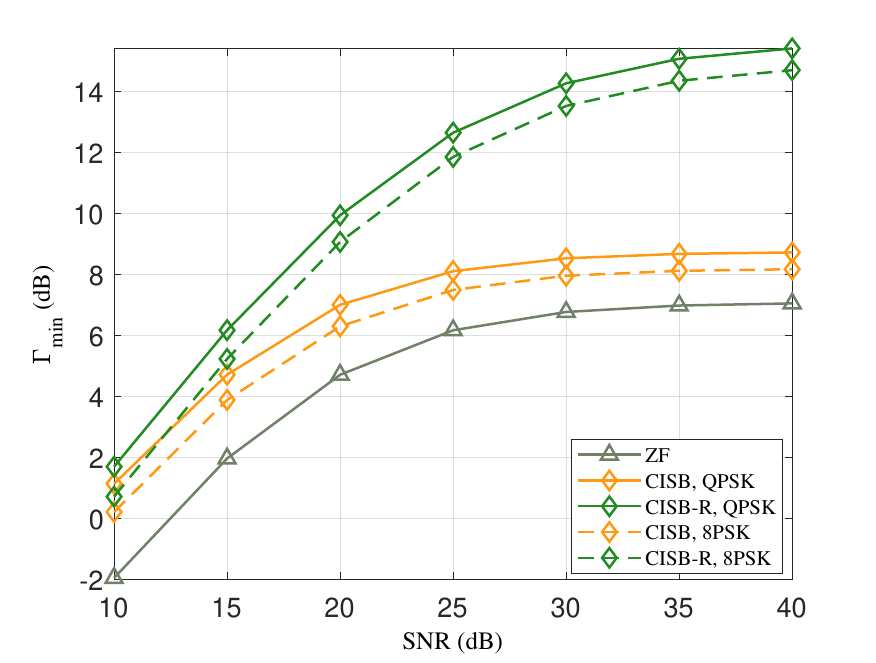}
		\caption{$\Gamma_{{\rm min}}$ vs SNR, ULA, $N=14$, $K=12$, $\alpha=0.995$.}
		\label{ULA Gamma}	
	\end{figure}
	
	\begin{figure}[htp]
		\centering
		\includegraphics[width=3.5in]{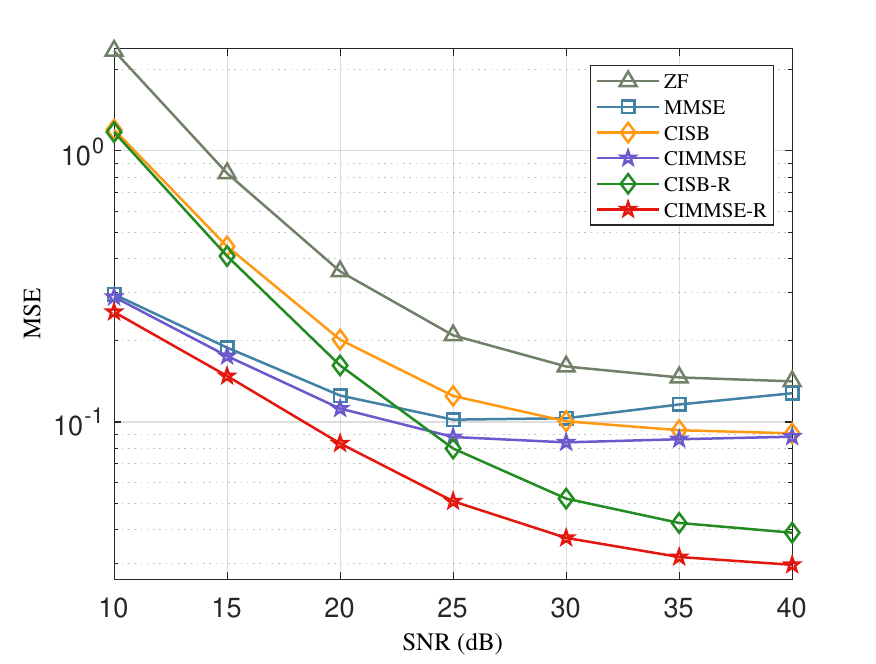}
		\caption{MSE vs SNR, ULA, $N=14$, $K=12$, $\alpha=0.995$, 8PSK.}
		\label{ULA MSE}	
	\end{figure}
	
	\begin{figure}[htb]
		\centering
		\includegraphics[width=3.5in]{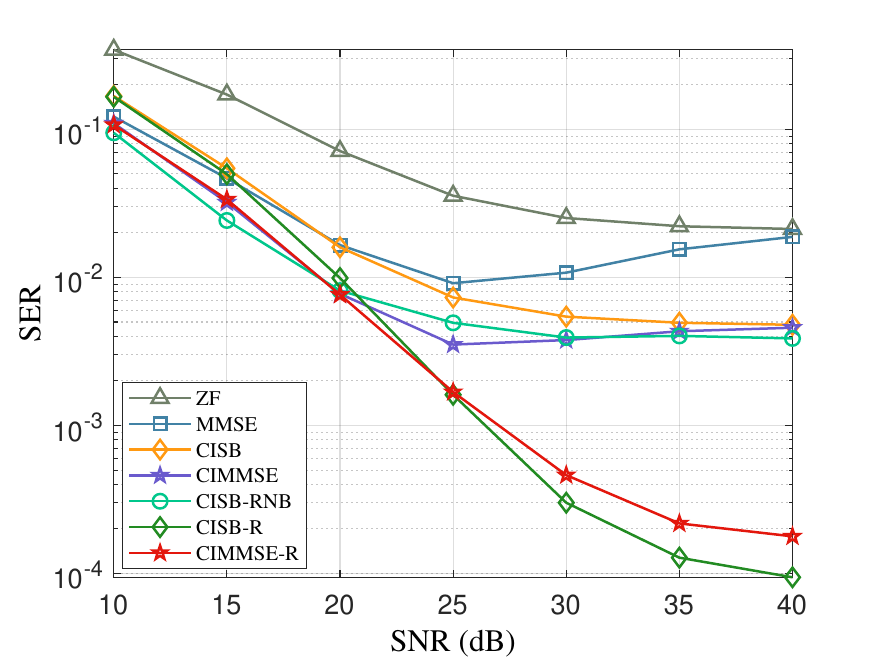}
		\caption{SER vs SNR, ULA, $N=14$, $K=12$, $\alpha=0.995$, QPSK.}
		\label{ULA SER}
	\end{figure}
	
	The comparison of $\Gamma_{\rm min}$, MSE, and SER performances for ULA with $N=14$, $K=12$, and $\alpha=0.995$ are shown in Fig. \ref{ULA Gamma}, \ref{ULA MSE}, and \ref{ULA SER}. As shown in Fig. \ref{ULA Gamma}, $\Gamma_{\rm min}$ of `CISB-R' is much higher than `CISB', and the gain becomes larger as the SNR increases, while gains of about 6.5dB can be observed when SNR is 40 dB. This indicates that our proposed scheme significantly improves the performance of the worst user.
	In Fig. \ref{ULA MSE}, although the MSEs of `CISB' and `CIMMSE' are smaller than `ZF' and `MMSE', they level off as the SNR increases due to the impact of imperfect CSI. In contrast, `CISB-R' and `CIMMSE-R' provide significantly lower MSE performance, which further decreases in high SNR regimes. Fig. \ref{ULA SER} shows that, for QPSK, the proposed two schemes achieve lower SER than other schemes, and they provide SNR gains of about 8dB than `CISB-RNB' when SER is $10^{-3}$ for QPSK. It is noteworthy that the performance of `MMSE' and `CIMMSE' even become worse in high SNR regimes since they respectively reduce to `ZF' and `CISB' as the SNR increases while ignoring the impact of imperfect CSI. Besides, all the above comparisons are applicable to the case of 8PSK.	

	\begin{figure}[htbp]
		\centering
		\includegraphics[width=3.5in]{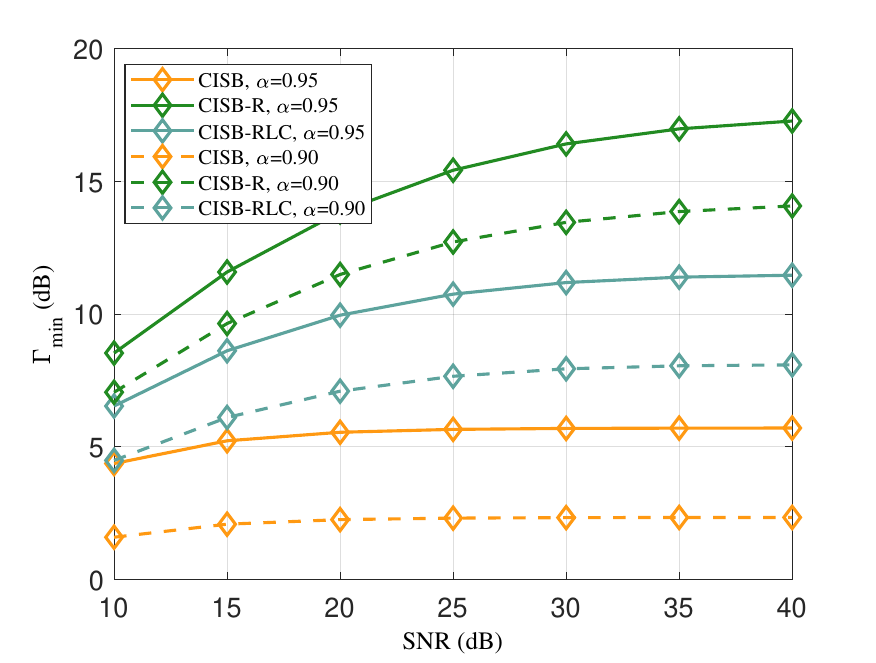}
		\caption{$\Gamma_{{\rm min}}$ vs SNR, UPA, $N=64$, $K=9$, 8PSK.}
		\label{UPA Gamma}
	\end{figure}
	\begin{figure}[htbp]
		\centering
		\subfigure[$\alpha=0.95$.]{
			\includegraphics[width=3.5in]{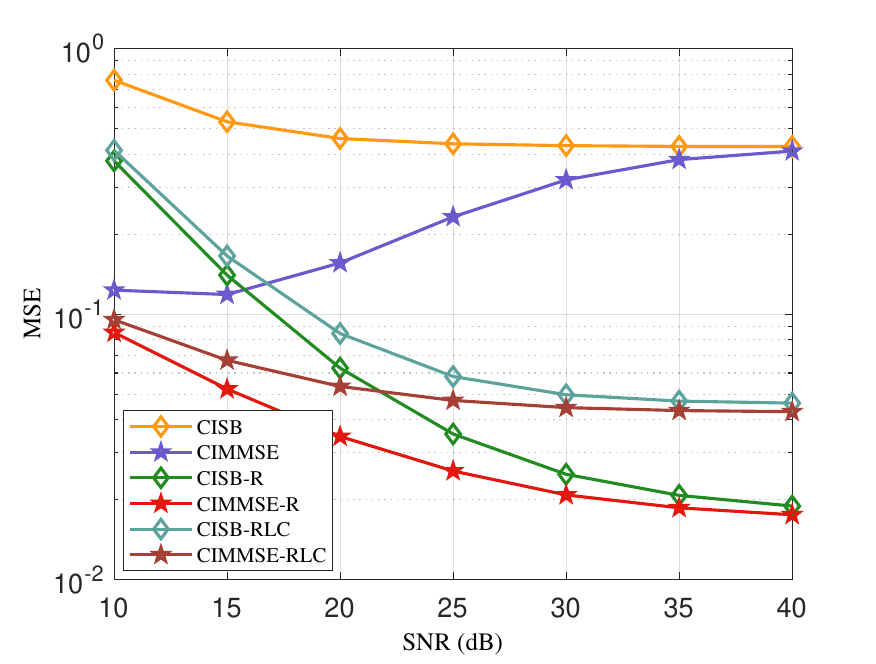}
		}
		\subfigure[$\alpha=0.90$.]{
			\includegraphics[width=3.5in]{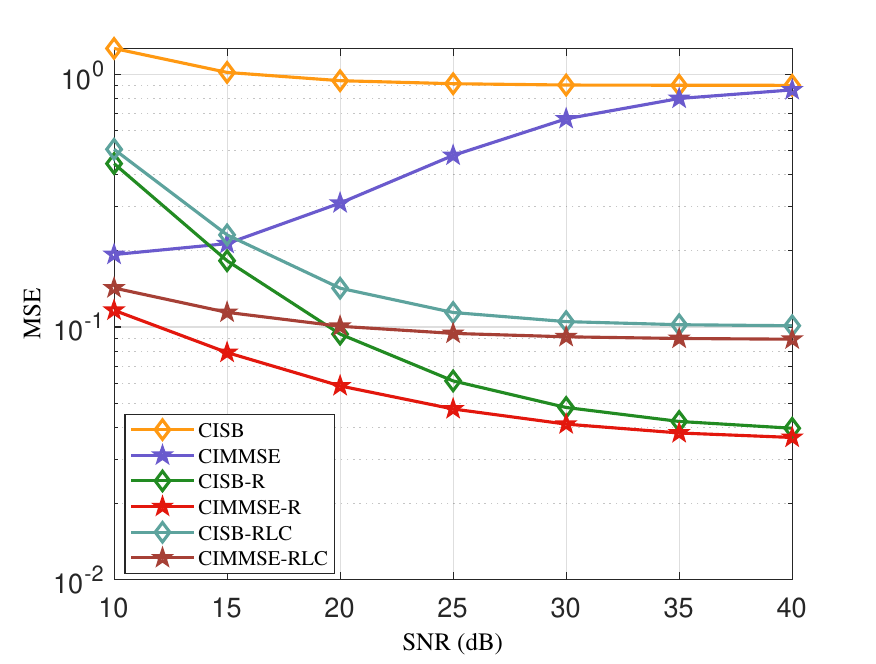}
		}
		\DeclareGraphicsExtensions.
		\caption{MSE vs SNR, UPA, $N=64$, $K=9$, 8PSK.}
		\label{UPA MSE}
	\end{figure}
	\begin{figure}[htbp]
		\centering
		\includegraphics[width=3.5in]{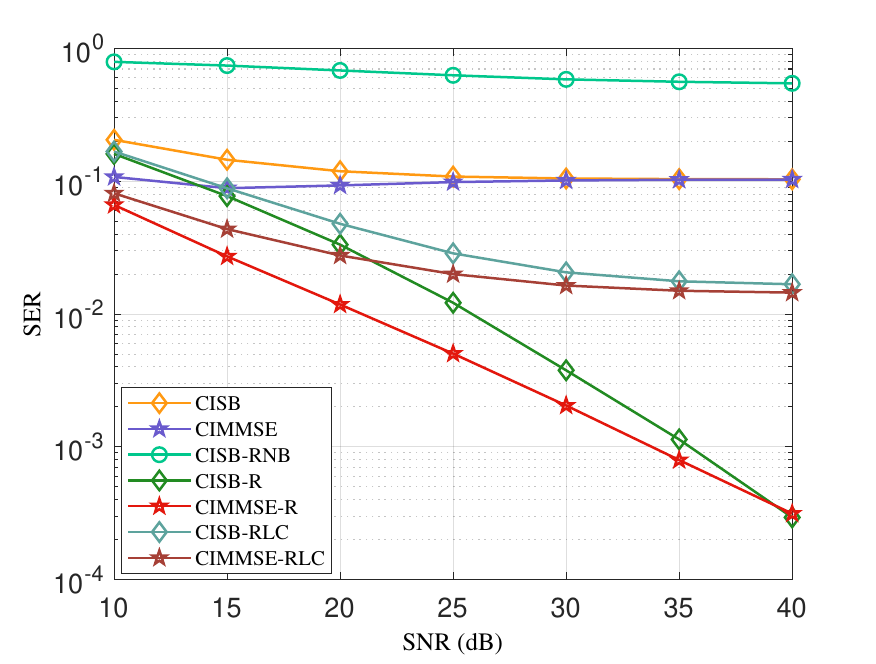}
		\caption{SER vs SNR, UPA, $N=64$, $K=9$, $\alpha=0.95$, 8PSK.}
		\label{UPA SER}
	\end{figure}

	The comparison of $\Gamma_{\rm min}$, MSE, and SER performances for UPA with $N=64$, $K=9$, and 8PSK are shown in Fig. \ref{UPA Gamma}, \ref{UPA MSE}, and \ref{UPA SER}. Since `CISB' and `CIMMSE' exhibit comparable performance to `ZF' and `MMSE' in this case, we omit the performance results of `ZF' and `MMSE'.
	 Compared to the performance in the case of ULA with $N\approx K$, `CISB-R' and `CIMMSE-R' provide more significant performance gains than `CISB' and `CIMMSE' in the case of UPA with $N\gg K$. Fig. \ref{UPA Gamma} shows that $\Gamma_{{\rm min}}$ of `CISB-R' and `CISB-RLC' are much higher than `CISB' given the same SNR, and our proposed two schemes even provide gains of about 11.6dB and 5.8dB when SNR is 40dB for $\alpha=0.95$, while the gains are almost same as the case of $\alpha=0.90$. As Fig. \ref{UPA MSE} shows, compared with other schemes whose MSE fluctuates $10^{-1}$ and $10^{0}$, the MSE of `CISB-R', `CIMMSE-R', `CISB-RLC' and `CIMMSE-RLC' approach the level of $10^{-2}$ with the increasing SNR. 
	 It can be observed from Fig. \ref{UPA SER} that the proposed schemes provide robust transmission with quite low SER, while the SER of other conventional schemes is larger than $10^{-1}$. Specifically, the SER of `CISB-RLC' and `CIMMSE-RLC' approach $10^{-2}$ in high SNR regime, and `CISB-R' and `CIMMSE-R' achieve better SER performance approaching $10^{-4}$. 
	 
	 {In \figref{UPA MSE} and \figref{UPA SER}, it can be observed that the MMSE-based schemes usually outperform those based on SINR balancing in terms of the average SER evaluation criterion. 
	 This implies that the strategy of MMSE-based methods focusing on average MSE has a greater impact on reducing average transmission SER.}
	 {`CIMMSE' performs worse in high SNR ranges since imperfect CSI introduces large interference, and the large regularization factor in `CIMMSE' in lower SNR ranges is more suitable for handling such interference \cite{1391204}.} Furthermore, `CISB-RNB' can hardly work in this scenario since the error bound $\delta^2_k$ is too large, which will result in the feasible domain of `CISB-RNB' being the empty set \cite{7103338}.
	 {Besides, the performance gaps between the proposed low-complexity schemes and the original ones gradually increase with the improvement of SNR, which is consistent with our analysis in Section \ref{LC robust SINR}.} {
	 Unlike the $\alpha=0.995$ used in the configuration with $N=14$, we consider $\alpha=0.95$ in the configuration with $N=64$. The smaller $\alpha$ (higher speed) was chosen because larger antenna arrays yield higher spatial diversity gains, allowing SLP to combat greater impact of imperfect CSI. Conversely, with $\alpha=0.95$, while the proposed schemes significantly outperform other schemes under the configuration with $N=14$, these results are not presented as the overall performance is unsatisfactory for transmission.}	
	
	\begin{figure*}[htbp]
		\centering
		\subfigure[SER vs $\alpha$.]{
			\begin{minipage}[t]{0.31\linewidth}
				\centering
				\includegraphics[width=1\textwidth]{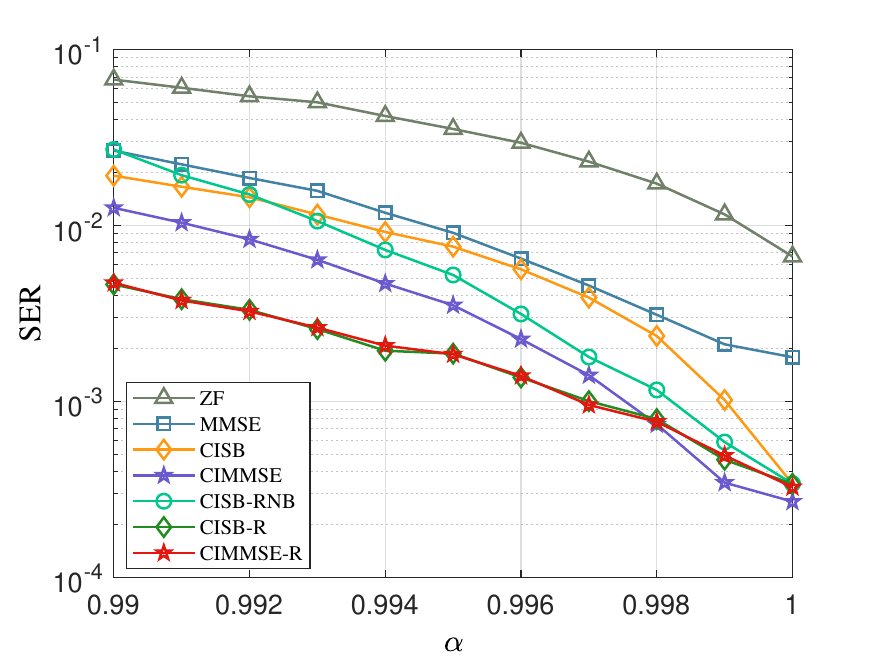}
			\end{minipage}%
		}%
		\subfigure[MSE vs $\alpha$.]{
			\begin{minipage}[t]{0.31\linewidth}
				\centering
				\includegraphics[width=1\textwidth]{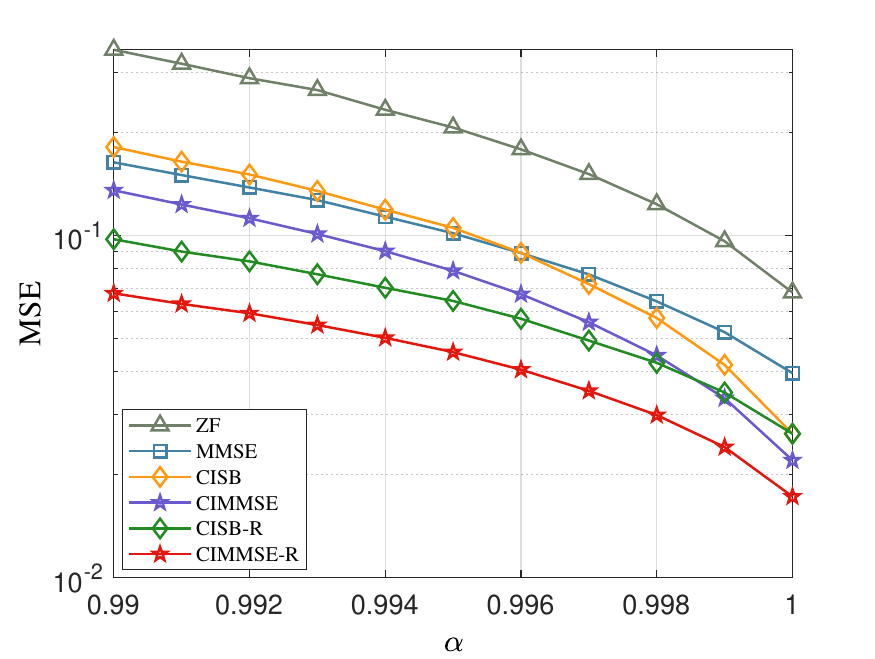}
			\end{minipage}
		}
		\subfigure[$\Gamma_{{\rm min}}$ vs $\alpha$.]{
			\begin{minipage}[t]{0.31\linewidth}
				\centering
				\includegraphics[width=1\textwidth]{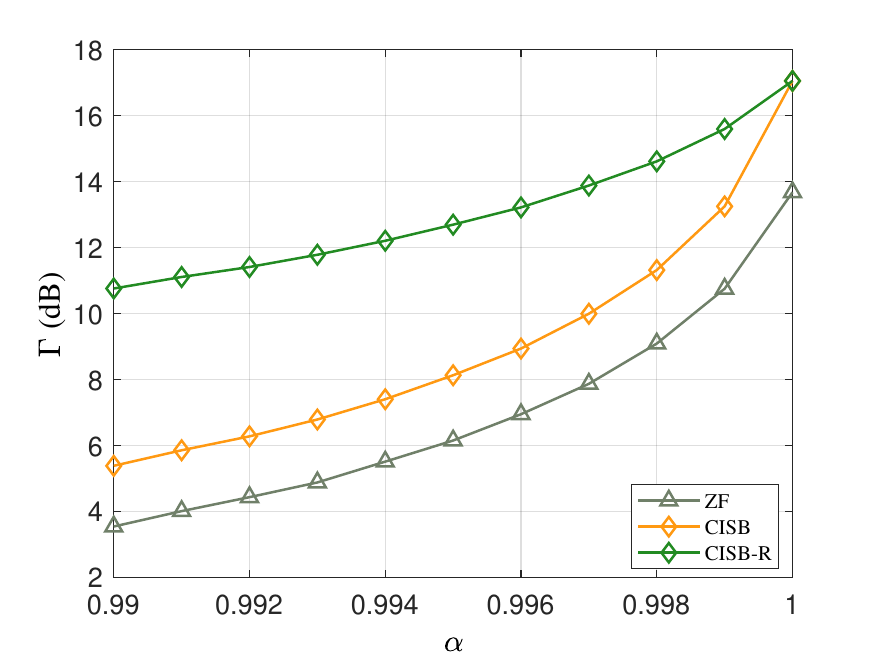}
			\end{minipage}
		}
		\centering
		\caption{{The performance with differernt $\alpha$, ULA, $N=14$, $K=12$, QPSK, SNR=25dB.}}
		\label{ULA multialpha}
	\end{figure*}
	\begin{figure*}[htbp]
		\centering
		\subfigure[SER vs $\alpha$.]{
			\begin{minipage}[t]{0.31\linewidth}
				\centering
				\includegraphics[width=1\textwidth]{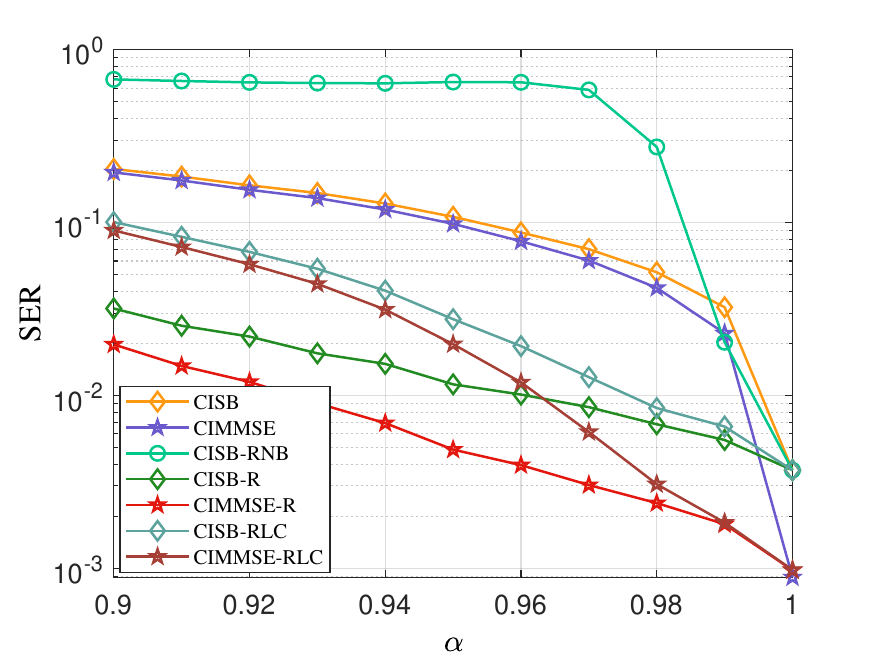}
			\end{minipage}%
		}%
		\subfigure[MSE vs $\alpha$.]{
			\begin{minipage}[t]{0.31\linewidth}
				\centering
				\includegraphics[width=1\textwidth]{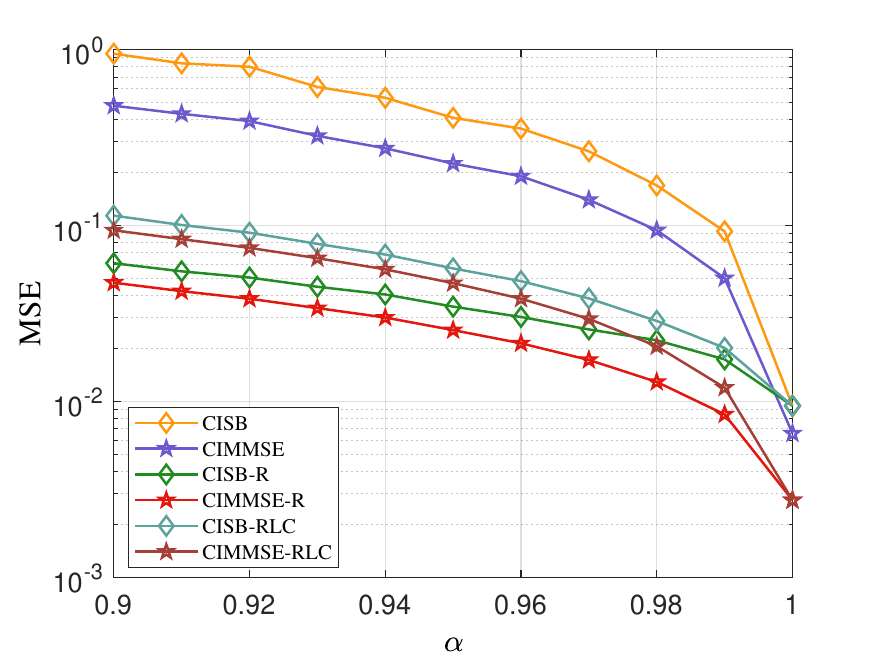}
			\end{minipage}
		}
		\subfigure[$\Gamma_{{\rm min}}$ vs $\alpha$.]{
			\begin{minipage}[t]{0.31\linewidth}
				\centering
				\includegraphics[width=1\textwidth]{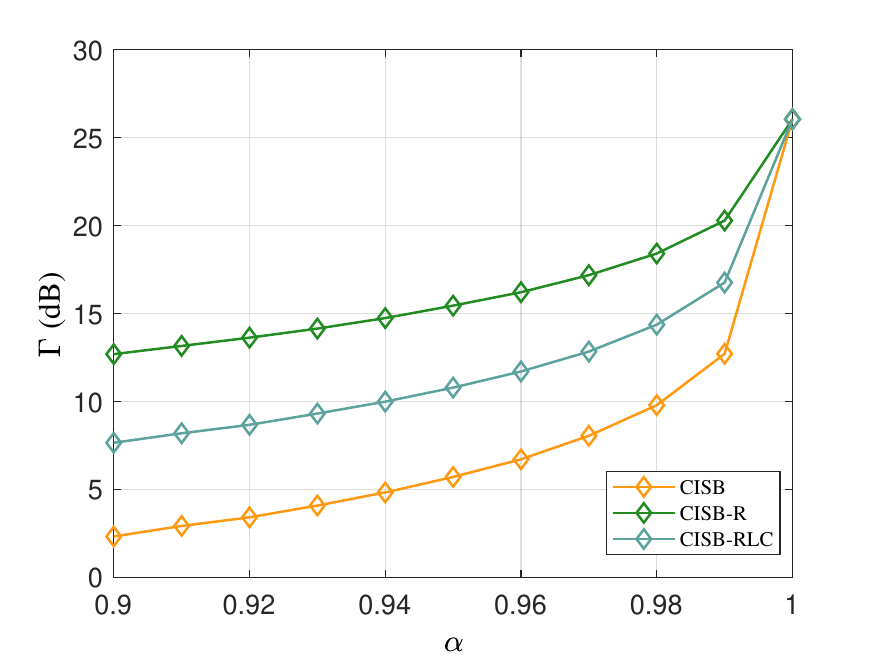}
			\end{minipage}
		}
		\centering
		\caption{{The performance with differernt $\alpha$, UPA, $N=64$, $K=9$, 8PSK, SNR=25dB.}}
		\label{UPA multialpha}
	\end{figure*}
	
	{In \figref{ULA multialpha} and \figref{UPA multialpha}, we compared the performance  under different $\alpha$ with two configurations, illustrating the performance comparison under varied UE speeds. It can be observed that our proposed scheme exhibits gains across nearly all considered $\alpha$ values. As $\alpha$ approaches 1, the impact of imperfect CSI diminishes, resulting in an overall performance improvement across all schemes, and the performance gaps between robust and non-robust approaches decrease. Under perfect CSI condition that $\alpha=1$, `CISB-RNB', `CISB-R', and `CISB-RLC' exhibit the same performance to `CISB', validating the conclusion in \remarkref{CISB remark} that the robust scheme degenerates to `CISB' under perfect CSI. In contrast to \remarkref{CIMMSE remark}, as we do not constrain $\gamma_1=\cdots=\gamma_K$, there exists a slight disparity between the performance of `CIMMSE-R' and `CIMMSE'. Furthermore, as shown in \figref{UPA multialpha}, the performance gaps between the proposed low-complexity schemes and the original ones gradually diminish with increasing $\alpha$, aligning with our analysis in Section \ref{LC robust SINR}.}
	
\begin{figure}[t]
		\centering
		\includegraphics[width=3.5in]{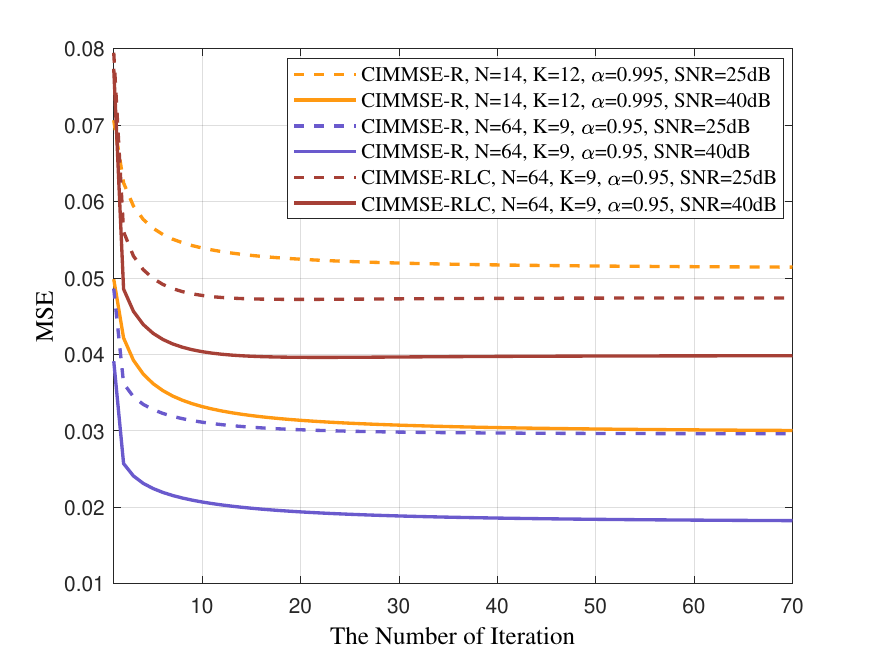}
		\caption{The convergence of \textbf{Algorithm 2} for 8PSK under different
			scenarios.}
		\label{convergence}
	\end{figure}

	Fig. \ref{convergence} shows the convergence of \algref{A4} for 8PSK. It can be observed that the objective function decreases as the
	number of iterations until convergence. The MSEs drop sharply in the first few iterations, which means `CIMMSE-R' and `CIMMSE-RLC' still work well if only a few iterations are performed in the considered scenario.
	
	\section{Conclusion}\label{Section 7}
	This paper focused on the design of robust SLP with imperfect CSI caused by channel aging.
	Starting from the widely adopted jointly correlated channel model, 
 	we considered the imperfect CSI as the statistical CSI with spatial correlation, which was utilized to formulate the signal model for the downlink SLP transmission with channel aging. On this basis, the SINR balancing and MSE minimization problems were formulated for robust SLP design, where the former targets to maximize the minimum SINR, and the latter aims to minimize the MSE between the received signal and the target constellation point. In the case of massive MIMO, we derived a closed-form SLP scheme for SINR balancing by approximating the objective function. Furthermore, an MMSE-based robust SLP with lower computational complexity was also developed by modifying the proposed algorithm. Simulation results, including comparisons of MSE, SINR, and SER, demonstrated the superiority of the proposed schemes. {In the future work, we plan to extend the proposed schemes to QAM transmission and employ hybrid precoding architectures. Additionally, we will consider combining the SINR balancing and MMSE problems to further improve the transmission performance.} 
		
	\appendices
	\section{Proof of Proposition \ref{constraint transformation}} \label{constraint transformation proof}
	{The first and third constraints in problem (17) are as follows:
			\begin{align}
			\begin{split}
			&\quad\text{s.t.} \;\; {{\bf H}}{{\bf x}}={{\boldsymbol{\Gamma}}}\left({{\bf s}}+{\boldsymbol{\Lambda}}{\boldsymbol{\delta}}\right),\\
			&\quad\quad\ \ {\boldsymbol{\delta}}\succeq {\bf 0}.
			\end{split}
			\label{two constraint}
			\end{align}
			${{\boldsymbol{\Gamma}}}$ can be decomposed into ${\bf I}_2\otimes {\bar{\boldsymbol{\Gamma}}}$, where ${\bar{\boldsymbol{\Gamma}}}={\rm diag}\{\gamma_1, ..., \gamma_K\}$. According to \cite{9910472}, ${\boldsymbol{\Lambda}}=\begin{bmatrix}
			{\bf M}_R & {\bf N}_R \\
			{\bf M}_I & {\bf N}_I
			\end{bmatrix},$ where ${\bf M}_R$, ${\bf N}_R$, ${\bf M}_I$, and ${\bf N}_I$ are four diagonal matrices. Then we have
			\begin{align}
			{{\boldsymbol{\Gamma}}}{\boldsymbol{\Lambda}} = 
			\begin{bmatrix}
			{\bar{\boldsymbol{\Gamma}}} & \\
		 & {\bar{\boldsymbol{\Gamma}}}
			\end{bmatrix}
			\begin{bmatrix}
			{\bf M}_R & {\bf N}_R \\
			{\bf M}_I & {\bf N}_I
			\end{bmatrix}
			= 
			\begin{bmatrix}
			{\bar{\boldsymbol{\Gamma}}}{\bf M}_R & {\bar{\boldsymbol{\Gamma}}}{\bf N}_R \\
			{\bar{\boldsymbol{\Gamma}}}{\bf M}_I & {\bar{\boldsymbol{\Gamma}}}{\bf N}_I
			\end{bmatrix}
			= {\boldsymbol{\Lambda}}{{\boldsymbol{\Gamma}}}.
			\end{align}
			Therefore, the first constraint in \eqref{two constraint} can be written as ${{\bf H}}{{\bf x}}-{{\boldsymbol{\Gamma}}}{{\bf s}}={\boldsymbol{\Lambda}}{{\boldsymbol{\Gamma}}}{\boldsymbol{\delta}}$. Since we constraint $\gamma_k>0$ and ${\boldsymbol{\delta}}\succeq {\bf 0}$, we define a new `${\boldsymbol{\delta}}$' as ${{\boldsymbol{\Gamma}}}{\boldsymbol{\delta}}$, which still satisfies ${\boldsymbol{\delta}}\succeq {\bf 0}$. Then, the constraint is transformed into ${{\bf H}}{{\bf x}}-{{\boldsymbol{\Gamma}}}{{\bf s}}={\boldsymbol{\Lambda}}{\boldsymbol{\delta}}$. }
	
	{Next, we prove that  ${\boldsymbol{\Lambda}}$ is an invertible matrix.
				According to \cite{9910472}, we have  
				\begin{align}
				\begin{split}
				{\bf M}_R &= {\rm diag}\{\real({{\mu}}_{1}),...,\real({{\mu}}_{K})\},\\
				{\bf M}_I &= {\rm diag}\{\imaginary({{\mu}}_{1}),...,\imaginary({{\mu}}_{K})\},\\
				{\bf N}_R &= {\rm diag}\{\real({{\nu}}_{1}),...,\real({{\nu}}_{K})\},\\ 
				{\bf N}_I &= {\rm diag}\{\imaginary({{\nu}}_{1}),...,\imaginary({{\nu}}_{K})\}, 
				\end{split}
				\label{M and N}
				\end{align}
				where ${{\mu}}_{k}$ and ${{\nu}}_{k}$ are two normal boundary parameters of the CIR belonging to $s_k$ and can be easily obtained from $s_k$. 
				Taking 8PSK in \figref{CI_MMSE_CIR} as an example, $s_k$ represents a PSK constellation point, with dashed lines on both sides denoting the decision boundary of this constellation point. The green area signifies the CIR for $s_k$, while the red lines, denoted by $\nu_k$ and $\mu_k$, represent the boundaries of the CIR, parallel to the decision boundaries of the constellation point. For the CIR of $M$-PSK constellation points ($M>2$), one of the two boundaries has a non-zero real part. We define $\mu_k$ as the boundary with a non-zero real part, while $\nu_k$ represents the other boundary. If both boundaries have a non-zero real part, either one can serve as $\mu_k$.
				Based on \eqref{M and N} and the aforementioned definition of $\mu_k$, ${\bf M}_R$ is an invertible diagonal matrix. Then, ${\boldsymbol{\Lambda}}$ is invertible if and only if its Schur complement ${\boldsymbol{\Lambda}}/{\bf M}_R={\bf N}_I-{\bf M}_I{\bf M}^{-1}_R{\bf N}_R$ is invertible \cite{zhang2006schur}. As matrix ${\boldsymbol{\Lambda}}/{\bf M}_R$ is diagonal, ${\boldsymbol{\Lambda}}$ is invertible when $\imaginary(\nu_k)-\imaginary(\mu_k)\real(\nu_k)/\real(\mu_k)\neq 0,\ \forall k\in{\mathcal K}$. We demonstrate the invertibility of ${\boldsymbol{\Lambda}}$ by showing its converse, that is, $\imaginary(\nu_k)-\imaginary(\mu_k)\real(\nu_k)/\real(\mu_k)=0$ does not hold. On the one hand, if $\real(\nu_k)\neq 0$, this equation implies $\frac{\imaginary(\nu_k)}{\real(\nu_k)}=\frac{\imaginary(\mu_k)}{\real(\mu_k)}$. This means that the two boundaries of the CIR are either in the same direction or in opposite directions, which does not hold for PSK constellation points when $M>2$. on the other hand, if $\real(\nu_k)= 0$, the equation implies $\imaginary(\nu_k)= 0$. As $\nu_k$ is non-zero, this does not hold. Thus, $\imaginary(\nu_k)-\imaginary(\mu_k)\real(\nu_k)/\real(\mu_k)=0$ does not hold, and ${\boldsymbol{\Lambda}}$ is invertible. On this basis, the first constraint is transformed from ${{\bf H}}{{\bf x}}-{{\boldsymbol{\Gamma}}}{{\bf s}}={\boldsymbol{\Lambda}}{\boldsymbol{\delta}}$ into ${\boldsymbol{\Lambda}}^{-1}({{\bf H}}{{\bf x}}-{{\boldsymbol{\Gamma}}}{{\bf s}})={\boldsymbol{\delta}}$. Combined with the constraint ${\boldsymbol{\delta}}\succeq{\bf 0}$, these two constraints can be merged into ${\boldsymbol{\Lambda}}^{-1}({{\bf H}}{{\bf x}}-{{\boldsymbol{\Gamma}}}{{\bf s}})\succeq {\bf 0}$. This concludes the proof.}
			
	\section{Proof of Proposition \ref{closed form T}} \label{closed form T proof}
	We force $\frac{\gamma_1}{\tau_1}=\frac{\gamma_2}{\tau_2}=\cdots=\frac{\gamma_K}{\tau_K} \triangleq{\tilde \gamma}$, based on which problem \eqref{robust SINR balancing LC1} can be converted into the following problem:
		\begin{align}
			\begin{split}
				&\max \limits_{{{\bf x}},{\boldsymbol{\delta}},{\tilde \gamma}} {\tilde \gamma}\\  
				&\quad\text{s.t.} \;\; {{\bf H}}{{\bf x}}={\tilde \gamma}{{\boldsymbol{\Theta}}}\left({{\bf s}}+{\boldsymbol{\Lambda}}{\boldsymbol{\delta}}\right),\\
				&\quad\quad\ \ \left\|{{\bf x}}\right\|_{2}^{2} = P_{\rm T},\\
				&\quad\quad\ \ {\boldsymbol{\delta}}\succeq {\bf 0}.
			\end{split}
			\label{key}
		\end{align}
		It can be verified that ${\bf x}^{\star}, {\boldsymbol{\delta}}^{\star}, {\tilde \gamma}^{\star}\tau_1, ..., {\tilde \gamma}^{\star}\tau_K$ is one of the optimal solutions to problem \eqref{robust SINR balancing LC1},
		where ${\bf x}^{\star}, {\boldsymbol{\delta}}^{\star}, {\tilde \gamma}^{\star}$ is the optimal solution to the above problem.
			 
		According to \cite{7042789}, the optimal solution to SINR balancing problem \eqref{key} can be easily obtained from that to the following PM problem:
		\begin{align}
			\begin{split}
				&\min\limits_{{{\bf x}},{\boldsymbol{\delta}}} \left\|{{\bf x}}\right\|_{2}^{2}\\  
				&\quad\text{s.t.} \;\; {{\bf H}}{{\bf x}}={{\boldsymbol{\Theta}}}\left({{\bf s}}+{\boldsymbol{\Lambda}}{\boldsymbol{\delta}}\right),\\
				&\quad\quad\ \ {\boldsymbol{\delta}}\succeq {\bf 0},
			\end{split}
			\label{approx PM}
		\end{align}
		where the optimal ${\boldsymbol{\delta}}^{\star}$ 
		can be obtained by solving the following NNLS problem:
		\begin{align}
			{\boldsymbol{\delta}}^{\star}=\arg\min\limits_{{\boldsymbol{\delta}}\succeq {\bf 0}}\|\sqrt{N}{{\bf H}}^{\dagger}{{\boldsymbol{\Theta}}}\left({{\bf s}}+{\boldsymbol{\Lambda}}{\boldsymbol{\delta}}\right)\|^2_2.
			\label{closed form NNLS}
		\end{align}
		Given ${\boldsymbol{\delta}}^{\star}$, the optimal ${{\bf x}}$ in \eqref{approx PM} is ${{\bf H}}^{\dagger}{{\boldsymbol{\Theta}}}\left({{\bf s}}+{\boldsymbol{\Lambda}}{\boldsymbol{\delta}}^{\star}\right)$ \cite{8815429}. By scaling the transmit signal ${\bf x}$ to the transmit power $P_{\rm T}$ \cite{7042789}, the optimal ${\tilde \gamma}^{\star}$ and ${{\bf x}}^{\star}$ of problem \eqref{key} are given by
		\begin{align}
		{\tilde \gamma}^{\star} &= \sqrt{\frac{P_{\rm T}}{\|{{\bf H}}^{\dagger}{{\boldsymbol{\Theta}}}\left({{\bf s}}+{\boldsymbol{\Lambda}}{\boldsymbol{\delta}}^{\star}\right)\|^2_2}},\label{optimal tilde gamma}\\
		{{\bf x}}^{\star} &= {\tilde \gamma}^{\star}{{\bf H}}^{\dagger}{{\boldsymbol{\Theta}}}\left({{\bf s}}+{\boldsymbol{\Lambda}}{\boldsymbol{\delta}}^{\star}\right).\label{optimal x}
		\end{align}
		Moreover, the corresponding optimal $\gamma^{\star}_k$ of problem \eqref{robust SINR balancing LC1} is given by
		\begin{align}
			\begin{split}
				\gamma^{\star}_k = {\tilde \gamma}^{\star}\cdot\tau_k, \;\forall k \in {\mathcal{ K}}.
				\label{optimal gamma}
			\end{split}
		\end{align}
		In conclusion, one of the optimal solutions to problem \eqref{robust SINR balancing LC1} is ${{\bf x}}^{\star}$, ${\boldsymbol{\delta}}^{\star}$, and $\gamma^{\star}_1,...,\gamma^{\star}_K$ in \eqref{optimal x}, \eqref{closed form NNLS}, and \eqref{optimal gamma}. This concludes the proof.
	\section{Proof of Lemma \ref{L1}}\label{Lemma 1 proof}
	The objective function of \eqref{a NNLS form} can be rewritten as 
	\begin{align}
		\begin{split}
			\|{\bf A}\left({{\bf s}}+{\boldsymbol{\Lambda}}{\boldsymbol{\delta}}\right)\|^2_2={{\bf s}}^T{\bf A}^T{\bf A}{{\bf s}}+f_{\rm nnls}({\boldsymbol{\delta}}),
		\end{split}
	\end{align}
	where $f_{\rm nnls}({\boldsymbol{\delta}})={\boldsymbol{\delta}}^T{\boldsymbol{\Lambda}}^T{\bf A}^T{\bf A}{\boldsymbol{\Lambda}}{\boldsymbol{\delta}}+
	2{{\bf s}}^T{\bf A}^T{\bf A}{\boldsymbol{\Lambda}}{\boldsymbol{\delta}}$. Without loss of generality, for $M$-PSK ($M>2$) and $M$-QAM with constellation points symmetrically placed around the imaginary and real axes, $[{\boldsymbol{\Lambda}}{\boldsymbol{\delta}}]_k$ and $[{{\bf s}}]_k$ have the same positivity or negativity when $[{\boldsymbol{\Lambda}}\boldsymbol{\delta}]_k>0, \forall k \in \{1,...,2K\}$. Therefore, function $f_{\rm nnls}$ can be expressed as
	\begin{align}
		\begin{split}
			f_{\rm nnls}({\boldsymbol{\omega}})&={{\bf s}}^T{\boldsymbol{\Omega}}{\bf A}^T{\bf A}{\boldsymbol{\Omega}}{{\bf s}}+
			2{{\bf s}}^T{\bf A}^T{\bf A}{\boldsymbol{\Omega}}{{\bf s}}\\
			&={{\bf s}}^T\left({\boldsymbol{\Omega}}+2{\bf I}\right){\bf A}^T{\bf A}{\boldsymbol{\Omega}}{{\bf s}},
		\end{split}
	\end{align}
	where ${\boldsymbol{\Omega}}={{\rm diag}\{{\boldsymbol{\omega}}\}}$ and ${\boldsymbol{\omega}}=[\omega_1, ..., \omega_{2K}]\succeq{\bf 0}$ satisfying ${\boldsymbol{\Omega}}{{\bf s}}={\boldsymbol{\Lambda}}{\boldsymbol{\delta}}$, i.e., $[\boldsymbol{\omega}]_k = [\boldsymbol{\Lambda}{\boldsymbol{\delta}}]_{k}/{[{{\bf s}}]_k},\ \forall k \in \{1,...,2K\}$. When ${\boldsymbol{\delta}}={\bf 0}$, we have  ${\boldsymbol{\omega}}={\bf 0}$ and $f_{\rm nnls}({\boldsymbol{\omega}})=0$, which means the existence of ${\boldsymbol{\omega}}\succeq {\bf 0}$ that satisfies $f_{\rm nnls}({\boldsymbol{\omega}})<0$ is equivalent to the condition that ${\boldsymbol{\delta}}={\bf 0}$ is not the the optimal solution for \eqref{a NNLS form}.
	Function $f_{\rm nnls}$ can be further written as 
	\begin{align}
		\begin{split}
			f_{\rm nnls}({\boldsymbol{\omega}}) &= \sum_{m=1}^{2K}\sum_{n=1}^{2K}(\omega_{m}+2)[{{\bf s}}]_{m}\cdot \omega_{n}[{{\bf s}}]_{n}\cdot {\bf a}^T_m{\bf a}_n,
		\end{split}
	\end{align}
	where ${\bf a}_m$ is the $m$-th column of ${\bf A}$. If ${\bf R}_{\bf A}$ is a diagonal matrix, we have 
	\begin{align}
		f_{\rm nnls}({\boldsymbol{\omega}}) \!=\! \sum_{m=1}^{2K}{\omega_{m}}(\omega_{m}+2)[{{\bf s}}]^2_{m} \|{\bf a}_m\|^2_2\geq 0,\ \forall{\boldsymbol{\omega}}\succeq {\bf 0},
	\end{align}
	which means ${\boldsymbol{\delta}}^{\star}= {\bf 0}$ achieves the optimum. This concludes the proof.
	
	\section{Proof of Proposition \ref{SB P}}\label{SB NNLS P proof}
	
	According to the definition of ${\bf R}_{\bf A}$ in \eqref{a NNLS form}, we have ${\bf A}=\sqrt{N}{{\bf H}}^T({{\bf H}}{{\bf H}}^T)^{-1}{{\boldsymbol{\Theta}}}$ and ${\bf R}_{\bf A}={{\boldsymbol{\Theta}}}({{\bf H}}{{\bf H}}^T/N)^{-1}{{\boldsymbol{\Theta}}}$ for NNLS problem \eqref{NNLS robust SB}. Without loss of generality, we consider the case of $F_{\rm vh}=1$ with the normalized channel whose power $\mathbb{E}\{{\bf h}^H_k{\bf h}_k\}=\|{\bf m}_{k}\|^2_2$ scales as $\mathcal{O}(N)$ \cite{5673745}. With the assumptions in \cite{8815518}, the following conclusion can be easily proved
	\begin{align}
		\left(\frac{{\bar{\bf h}_{k_1}}^{H}{\bar{\bf h}_{k_2}}}{N}\right)_{N\gg K}{\overset{p}{\rightarrow}}
		\begin{cases}
			\frac{{\alpha}^2_{k_1}\|{\bf m}_{k_1}\|^2_2}{N},\ k_1=k_2,\ k_1,k_2\in{\mathcal K}\\
			0, \ k_1\neq k_2,\ k_1,k_2\in{\mathcal K}\\
		\end{cases}
		,
		\label{orthogonality 1}
	\end{align}
	where ${\overset{p}{\rightarrow}}$ denotes converging in probability with increasing $N$. The above formulation is also consistent with the asymptotic inter-terminal channel orthogonality \cite{6375940, 7166308}. We further have 
	\begin{align}
		\left(\frac{{{\bf H}}{{\bf H}}^T}{N}\right)_{N\gg K}{\overset{p}{\rightarrow}} {\bf I}_2\otimes {\bf M},
	\end{align}
	where ${\bf M}={\rm diag}\left\{\frac{{\alpha}^2_{1}\|{\bf m}_{1}\|^2_2}{N},...,\frac{{\alpha}^2_{K}\|{\bf m}_{K}\|^2_2}{N}\right\}$. According to the \textit{Continuous Mapping Theorem} \cite{van2000asymptotic}, it can be proved that ${\bf R}_{\bf A}$ converges to a diagonal matrix given by
	\begin{align}
		({\bf R}_{\bf A})_{N\gg K}{\overset{p}{\rightarrow}}({\bf I}_2\otimes{\bf M}^{-1}){{\boldsymbol{\Theta}}}^2,
		\label{orthogonality 2}
	\end{align}
	where the diagonal elements of ${{\boldsymbol{\Theta}}}^2$ scale as $\mathcal{O}(1)$. This concludes the proof, and the diagonal property of ${\bf R}_{\bf A}$ is also corroborated by Fig. \ref{AMat} (a).

	\section{Proof of Proposition \ref{CIMMSE P}}\label{Proposition 2 proof}
	Given certain $\boldsymbol{\delta}$, the problem \eqref{robust MMSE SLP sub1} degenerates into
	\begin{align}
		\begin{split}
			&\min_{{{\bf x}}, \eta}\  {\mathbb{E}}_{{{\bf n}}}\left\{\left\|\frac{{{\boldsymbol{\Psi}}} \left({{{\bf H}}{{\bf x}}+{{\bf n}}}\right)}{\eta}-{\tilde{\bf s}}\right\|^2_2\right\}\\
			&\quad{\rm s.t.}\ \left\|{{\bf x}}\right\|^2_2 = P_{\rm T},
		\end{split}
		\label{robust MMSE SLP sub1 sub}
	\end{align}	
	where the constraint $\eta>0$ is omitted temporarily. Without loss of optimality, the constraint $\left\|{{\bf x}}\right\|^2_2 \leq P_{\rm T}$ is replaced by $\left\|{{\bf x}}\right\|^2_2 = P_{\rm T}$ since it is readily to prove that the optimal ${\bf x}$ satisfies the latter constraint. The objective function of the above problem can be expressed as
	\begin{align}
		\begin{split}
			f_{\rm p1}&={\mathbb{E}}_{{{\bf n}}}\left\{\left\|\frac{{{\boldsymbol{\Psi}}} \left({{{\bf H}}{{\bf x}}+{{\bf n}}}\right)}{\eta}-{\tilde{\bf s}}\right\|^2_2\right\}\\
			&=
			\frac{\left\|{{\boldsymbol{\Psi}}}{{\bf H}}{{\bf x}}\right\|^2_2}{\eta^2}-\frac{2{{\tilde{\bf s}}^T{{\boldsymbol{\Psi}}}{{\bf H}}{{\bf x}}}}{\eta}+\|{\tilde{\bf s}}\|^2_2+{\mathbb{E}}_{{{\bf n}}}\left\{\frac{\|{{\boldsymbol{\Psi}}}{{\bf n}}\|^2_2}{\eta^2}\right\}.
		\end{split}
		\label{f sub1}
	\end{align}
	Since ${\bar{n}}_k\sim\mathcal{C}\mathcal{N}(0, {\beta}^2_k\|{{\bf V}}_k{{\bf x}}\|_{2}^{2}+\sigma^2)$, we have
	\begin{align}
		{\mathbb{E}}_{{{\bf n}}}\left\{\frac{\|{{\boldsymbol{\Psi}}}{{\bf n}}\|^2_2}{\eta^2}\right\} = \frac{\sum^{K}_{k=1}{\psi^2_k}\left({\beta}^2_k{{\bf x}}^T{\bf V}_{k}^T{\bf V}_{k}{{\bf x}}+\sigma^2\right)}{\eta^2}.
	\end{align}
	The Lagrangian function of \eqref{robust MMSE SLP sub1 sub} is given by
	\begin{align}
		L\left({{\bf x}}, \eta, \lambda\right) = f_{\rm p1} -\lambda\left[\left\|{{\bf x}}\right\|^2_2-P_{\rm T}\right],
	\end{align}
	whose derivations with respect to ${{\bf x}}^{\star}$, $\eta^{\star}$, and $\lambda^{\ast}$ must vanish. The derivation with respect to ${{\bf x}}$ is
	\begin{align}
		\frac{\partial L\left({{\bf x}}, \gamma, \lambda\right)}{\partial {{\bf x}}} = 2\frac{\left({{\bf H}}^T{{\boldsymbol{\Psi}}}^2{{\bf H}}+{\boldsymbol {\Upsilon}}\right){{\bf x}}}{\eta^2}-2\frac{{{\bf H}}^T{{\boldsymbol{\Psi}}}{\tilde{\bf s}}}{\eta}-2\lambda{{\bf x}}.
	\end{align}
	By vanishing the derivation, we have the expression of optimal ${{\bf x}}$:
	\begin{align}
		{{\bf x}}^{\star} = \eta\left({{\bf H}}^T{{\boldsymbol{\Psi}}}^2{{\bf H}}+{\boldsymbol {\Upsilon}}-\lambda\eta^2{\bf I}_{2N}\right)^{-1}{{\bf H}}^T{{\boldsymbol{\Psi}}}{\tilde{\bf s}}.
	\end{align}
	We further replace $\lambda\eta^2$ with $\xi$, and the above formula can be rewritten as 
	\begin{align}
		&{{\bf x}}\left(\xi\right) = \eta(\xi){\bf u}\left(\xi\right),\label{xi 1}\\
		&{\bf u}\left(\xi\right) = \left({{\bf H}}^T{{\boldsymbol{\Psi}}}^2{{\bf H}}+{\boldsymbol {\Upsilon}}+\xi{\bf I}_{2N}\right)^{-1}{{\bf H}}^T{{\boldsymbol{\Psi}}}{\tilde{\bf s}}.\label{xi 2}
	\end{align}
	Due to $\left\|{{\bf x}}\right\|^2_2 = P_{\rm T}$, we can represent $\eta$ by 
	\begin{align}
		\eta(\xi) = \sqrt{\frac{P_{\rm T}}{{\bf u}\left(\xi\right)^T{\bf u}\left(\xi\right)}}.\label{xi 3}
	\end{align}
	By replacing ${{\bf x}}$ and $\eta$ by ${{\bf x}}(\xi)$ and $\eta(\xi)$, the problem \eqref{robust MMSE SLP sub1 sub} is transformed into the following unconstrained problem
	\begin{align}
		\arg\min_{\xi}f_{\rm p1}({{\bf x}}(\xi), \gamma(\xi)).
	\end{align}
	Based on \eqref{xi 1}, \eqref{xi 2}, and \eqref{xi 3}, the optimal $\xi^{\star}$ can be found by vanishing the derivation with respect to it and is given by
	\begin{align}
		\xi^{\star} = \frac{\sigma^2\sum_{k=1}^{K}\psi^2_k}{P_{\rm T}}.
	\end{align}
	Thus, there exists optimal solution $({{\bf x}}^{\star}, \eta^{\star})$ of problem \eqref{robust MMSE SLP sub1} given by
	\begin{align}
		\begin{split}
			&{{\bf x}}^{\star}=\eta^{\star}\left({{\bf H}}^T{{\boldsymbol{\Psi}}}^2{{\bf H}}+{\boldsymbol {\Upsilon}}+\frac{{\sigma^2}\sum_{k=1}^{K}{\psi^2_k}}{P_{\rm T}}{\bf I}_{2N}\right)^{-1}{{\bf H}}^T{{\boldsymbol{\Psi}}}{\tilde{\bf s}},\\
			&\eta^{\star} = \sqrt{\frac{P_{\rm T}}{\left\|\left({{\bf H}}^T{{\boldsymbol{\Psi}}}^2{{\bf H}}+{\boldsymbol {\Upsilon}}+\frac{{\sigma^2}\sum_{k=1}^{K}{\psi^2_k}}{P_{\rm T}}{\bf I}_{2N}\right)^{-1}{{\bf H}}^T{{\boldsymbol{\Psi}}}{\tilde{\bf s}}\right\|^2_2}}.
		\end{split}
	\end{align}
	Since $\eta^{\star}>0$, omitting $\eta>0$ in \eqref{robust MMSE SLP sub1 sub} does not lose the optimality. This concludes the proof.
	
	\section{Proof of Proposition \ref{CIMMSE NNLS P}}\label{CIMMSE NNLS P proof}
\begin{figure}[htbp]
		\centering
		\includegraphics[width=3.5in]{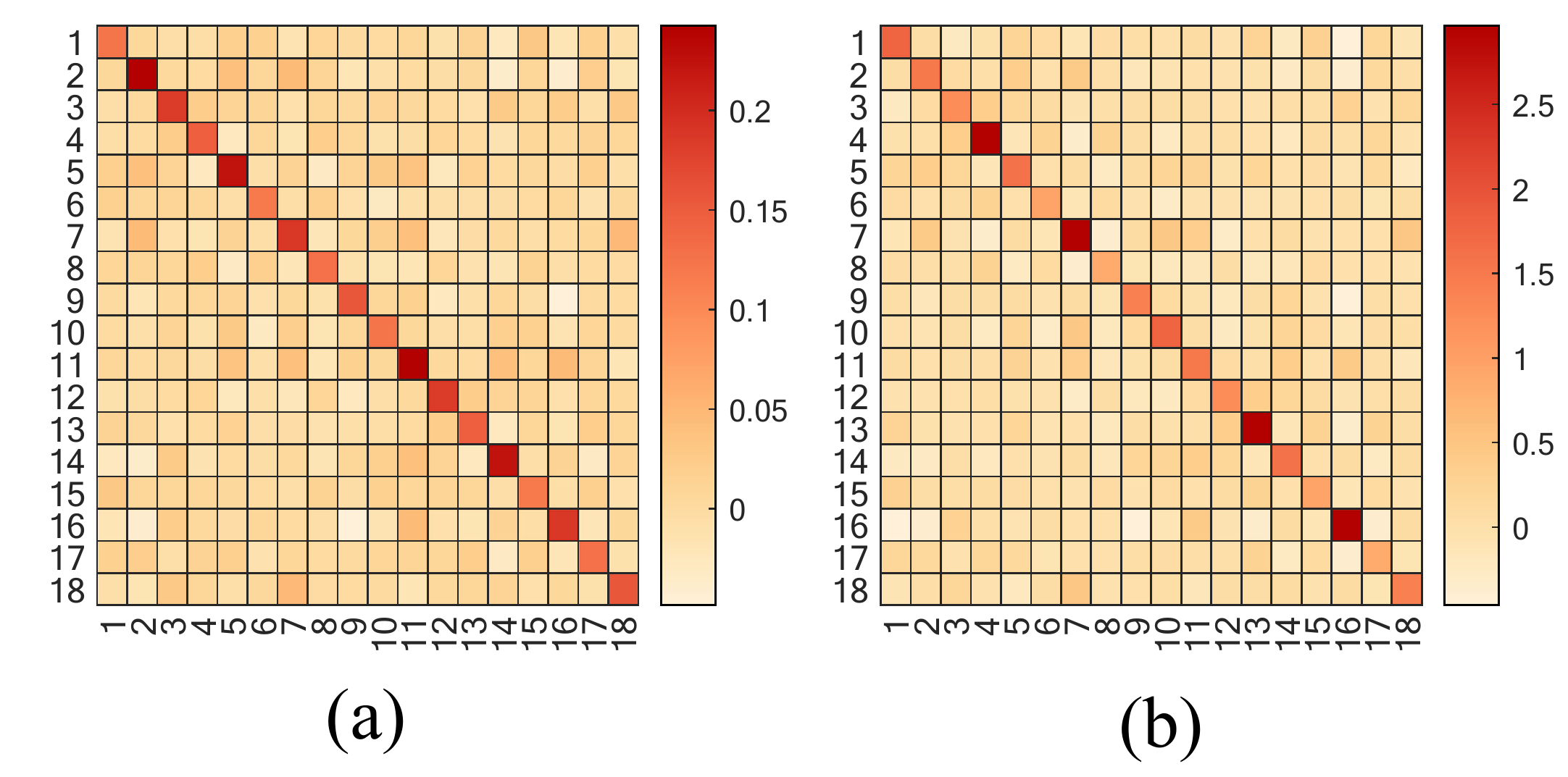}
		\caption{Heat map for the element magnitude of ${\bf R}_{\bf A}$ from (a) NNLS problem \eqref{NNLS robust SB} and (b) NNLS problem \eqref{robust MMSE NNLS}, $N=64$, $K=9$, $\alpha=0.95$, ${\rm SNR}=40$dB, 8PSK. The rest configuration is the same as the UPA in Section \ref{Section 6}.}
		\label{AMat}
	\end{figure}

	According to the definition in Appendix \ref{SB NNLS P proof}, matrix ${\bf R}_{\bf A}$ for problem \eqref{robust MMSE NNLS} is given by
	\begin{align}
		\begin{split}
			&{\bf R}_{\bf A}\\
			&=N({\bf I}_{2K}-{{\boldsymbol{\Psi}}}{{\bf H}}{\bf P})\\
			&=N\left[{\bf I}_{2K}-{{\boldsymbol{\Psi}}}{{\bf H}}\left({{\bf H}}^T{{\boldsymbol{\Psi}}}^2{{\bf H}}+\kappa{\bf I}_{2N}\right)^{-1}{{\bf H}}^T{{\boldsymbol{\Psi}}}\right]\\
			&=N\left[{{\boldsymbol{\Psi}}}{{\bf H}}\left(\kappa{\bf I}_{2N}\right)^{-1}{{\bf H}}^T{{\boldsymbol{\Psi}}}\right]^{-1}\\
			&=\kappa\left[{{\boldsymbol{\Psi}}}\left(\frac{{{\bf H}}{{\bf H}}^T}{N}\right){{\boldsymbol{\Psi}}}\right]^{-1}.
		\end{split}
	\end{align}
	Similar to \eqref{orthogonality 2}, it can be proved that ${\bf R}_{\bf A}$ converges to the diagonal matrix given by
	\begin{align}
		\left({\bf R}_{\bf A}\right)_{N\gg K} {\overset{p}{\rightarrow}} \kappa\left[{{\boldsymbol{\Psi}}}^2({\bf I}_2\otimes{\bf M})\right]^{-1},
	\end{align}
	where $\kappa$ scales as $\mathcal{O}(1)$. This concludes the proof, and
	the diagonal property of ${\bf R}_{\bf A}$ is also corroborated by Fig. \ref{AMat} (b).
	
	\ifCLASSOPTIONcaptionsoff
	\newpage
	\fi

	\bibliographystyle{IEEEtran}
	\bibliography{IEEEfull}
\end{document}